\documentstyle[12pt,epsfig,axodraw]{article}
\begin{document}
\pagestyle{plain} \setcounter{page}{1} \baselineskip=0.3in
\begin{titlepage}
\begin{flushright}
PKU-TH-2001-68
\end{flushright}
\vspace{.5cm}

\begin{center}
{\Large Supersymmetric Electroweak Corrections to Sbottom Decay
into Lighter Stop and Charged Higgs Boson }

\vspace{.2in}
Li Gang Jin and Chong Sheng Li  \\

\vspace{.2in}

Department of Physics, Peking University, Beijing 100871, \\
People's Republic of China
\\
\end{center}
\vspace{.4in}
\begin{footnotesize}
\begin{center}\begin{minipage}{5in}
\baselineskip=0.25in
\begin{center} ABSTRACT \end{center}
The Yukawa corrections of order ${\cal
O}(\alpha_{ew}m_{t(b)}^2/m_W^2)$, ${\cal
O}(\alpha_{ew}m_{t(b)}^3/m_W^3)$ and ${\cal
O}(\alpha_{ew}m_{t(b)}^4/m_W^4)$ to the width of sbottom decay
into lighter stop plus charged Higgs boson are calculated in the
Minimal Supersymmetric Standard Model. These corrections depend on
the masses of charged Higgs boson and lighter stop, and the
parameters $\tan\beta$ and $\mu$. For favorable parameter values,
the corrections decrease or increase the decay widths
significantly. Especially for high values of $\tan\beta$(=30) the
corrections exceed at least $10\%$ for both $\tilde b_1$ and
$\tilde b_2$ decay. But for low values of $\tan\beta$(=4,10) the
corrections are small and the magnitudes are less than $10\%$. The
numerical calculations also show that using the running bottom
quark mass which includes the QCD effects and resums all high
order $\tan\beta$-enhanced effects can improve much the
convergence of the perturbation expansion.

\end{minipage}\end{center}
\end{footnotesize}
\vfill

PACS number(s): 14.80.Cp; 14.80.Ly; 12.38.Bx

\end{titlepage}

\eject \baselineskip=0.3in
\section{Introduction}

The Minimal Supersymmetric Standard Model (MSSM)\cite{MSSM,Haber}
is an attractive extension of the Standard Model (SM). In this
Model every quark has two spin zero partners called squarks
$\tilde {q}_L$ and $\tilde {q}_R$, one for each chirality
eigenstate. These current eigenstates mix to form the mass
eigenstates $\tilde {q}_1$ and $\tilde {q}_2$. The third
generation squarks are of special interest. This is mainly due to
the reasons: large Yukawa couplings lead to strong mixing which
induces large mass differences between the lighter mass eigenstate
and the heavier one. This implies in general a very complex decay
pattern of the heavier states. The dominate decay modes of the
heavier squarks are the decays into quarks plus
charginos/neutralinos, decays into lighter squarks plus vector
bosons and decays into lighter squarks plus Higgs bosons. All
these squark decays have been extensively discussed at the
tree-level\cite{treelevel,Bartl,Hidaka}. The next generation of
colliders, for example, CERN Large Hadron Collider (LHC) will be
able to produce such kind of particles with masses up to 2.5
TeV\cite{LHC}, and a $e^+e^-$ linear Collider (LC)\cite{LC} will
be able to make precise measurements of their properties. Thus a
more accurate calculation of the decay mechanisms beyond the
tree-level is necessary to provide a solid basis for experimental
analysis of observing these decays at the next generation of
colliders. Up to now one-loop QCD and supersymmetric (SUSY) QCD
corrections to the squark decays have been
calculated\cite{Bartl,QCDcorrection}, and the Yukawa corrections
to the squark decays into quarks plus charginos/neutralinos also
were given in Ref.\cite{Yukawacorrection}. Very recently, a
complete one-loop computation of the electroweak radiative
corrections to the above processes has been presented by J.
Guasch, W. Hollik and J. Sol$\grave{\rm a}$\cite{full}. But the
electroweak radiative corrections to the heavier squark decays
into lighter squarks plus vector bosons and decays into lighter
squarks plus Higgs bosons have not been calculated yet, even
Yukawa corrections to these processes. In this paper, we present
the calculations of the Yukawa corrections of order ${\cal
O}(\alpha_{ew}m_{t(b)}^2/m_W^2)$, ${\cal
O}(\alpha_{ew}m_{t(b)}^3/m_W^3)$ and ${\cal
O}(\alpha_{ew}m_{t(b)}^4/m_W^4)$ to the width of sbottom decay
into lighter stop plus charged Higgs boson, i.e. the decay $\tilde
{b}_i \rightarrow \tilde {t}_1+ H^-$, where $\tilde {t}_1$ is the
lighter stop. These corrections are mainly induced by Yukawa
couplings from Higgs-quark-quark couplings, Higgs-squark-squark
couplings, Higgs-Higgs-squark-squark couplings,
chargino(neutralino)-quark-squark couplings, and
squark-squark-squark-squark couplings.

Our results can be generalized straightforwardly to the decay
$\tilde {t}_2\rightarrow \tilde {t}_1+(h^0,A^0)$. As for the
heavier squark decays into lighter squarks plus vector bosons, the
electroweak radiative corrections are simple because of the
relatively less renormalization parameters involved.

\section{Notation and tree-level result}

In order to make this paper self-contained, we first summarize our
notation and present the relevant interaction Lagrangians of the
MSSM and the tree-level decay rates for $\tilde {b}_i \rightarrow
\tilde {t}_1+ H^-$.

The current eigenstates $\tilde{q}_L$ and $\tilde{q}_R$ are
related to the mass eigenstates $\tilde{q}_1$ and $\tilde{q}_2$
by:
\begin{equation}
\left(\begin{array}{c} \tilde{q}_1 \\ \tilde{q}_2 \end{array}
\right)= R^{\tilde{q}}\left(\begin{array}{c} \tilde{q}_L \\
\tilde{q}_R \end{array} \right), \ \ \ \ \
R^{\tilde{q}}=\left(\begin{array}{cc} \cos\theta_{\tilde{q}} &
\sin\theta_{\tilde{q}} \\ -\sin\theta_{\tilde{q}} &
\cos\theta_{\tilde{q}}
\end{array} \right)
\end{equation}
with $0 \leq \theta_{\tilde{q}} < \pi$ by convention.
Correspondingly, the mass eigenvalues $m_{\tilde{q}_1}$ and
$m_{\tilde{q}_2}$ (with $m_{\tilde{q}_1}\leq m_{\tilde{q}_2}$) are
given by
\begin{eqnarray}\label{Mq2}
\left(\begin{array}{cc} m_{\tilde{q}_1}^2 & 0 \\ 0 &
m_{\tilde{q}_2}^2 \end{array} \right)=R^{\tilde{q}}
M_{\tilde{q}}^2 (R^{\tilde{q}})^\dag, \ \ \ \ \
M_{\tilde{q}}^2=\left(\begin{array}{cc} m_{\tilde{q}_L}^2 & a_qm_q
\\ a_qm_q & m_{\tilde{q}_R}^2 \end{array} \right)
\end{eqnarray}
with
\begin{eqnarray}
m^2_{\tilde{q}_L} &=& M^2_{\tilde{Q}} +m_q^2
+m_Z^2\cos2\beta(I_{3L}^q -e_q\sin^2\theta_W), \\
m^2_{\tilde{q}_R} &=& M^2_{\{\tilde{U},\tilde{D}\}} +m_q^2
+m_Z^2\cos2\beta e_q\sin^2\theta_W, \\
a_q &=& A_q -\mu\{\cot\beta, \tan\beta\}
\end{eqnarray}
for \{up,down\} type squarks. Here $M_{\tilde{q}}^2$ is the squark
mass matrix. $M_{\tilde{Q},\tilde{U},\tilde{D}}$ and $A_{t,b}$ are
soft supersymmetric breaking parameters and $\mu$ is the Higgs
mixing parameter in the superpotential. $I_{3L}^q$ and $e_q$ are
the third component of the weak isospin and the electric charge of
the quark $q$, respectively.

Defining $H_k=(h^0, H^0, A^0, G^0, H^{\pm}, G^{\pm})$ ($k$=1...6),
one can write the relevant lagrangian density in the
($\tilde{q}_1,\tilde{q}_2$) basis as following form ($i,j$=1,2;
$\alpha$ and $\beta$ flavor indices)
\begin{eqnarray}\label{lagrangian}
&& {\cal L}_{\rm relevant}=H_k\bar{q}^{\beta}(a_k^{\alpha}P_L
+b_k^{\alpha}P_R)q^{\alpha}
+(G^{\tilde{\alpha}}_k)_{ij}H_k\tilde{q}_j^{\beta\ast}\tilde{q}^\alpha_i
+g\bar{q}(a_{ik}^{\tilde{q}}P_R
+b_{ik}^{\tilde{q}})\tilde{\chi}^0_k\tilde{q}_i \nonumber \\
&& \hspace{1.5cm} +g\bar{t}(l_{ik}^{\tilde{b}}P_R
+k_{ik}^{\tilde{b}}P_L)\tilde{\chi}^+_k\tilde{b}_i
+g\bar{b}(l_{ik}^{\tilde{t}}P_R
+k_{ik}^{\tilde{t}}P_L)\tilde{\chi}_k^{+c}\tilde{t}_i \nonumber \\
&& \hspace{1.5cm} +(G^{\tilde{\alpha}}_{5k})_{ij}H^+H_k
\tilde{q}_j^{\beta\ast}\tilde{q}^\alpha_i +h.c.,
\end{eqnarray}
with
\begin{eqnarray}
(G^{\tilde{\alpha}}_k)_{ij} =[R^{\tilde{\alpha}}
\hat{G}_k^{\tilde{\alpha}} (R^{\tilde{\beta}})^T]_{ij},
\hspace{1.0cm} (G^{\tilde{\alpha}}_{5k})_{ij} =[R^{\tilde{\alpha}}
\hat{G}_{5k}^{\tilde{\alpha}} (R^{\tilde{\beta}})^T]_{ij}
\hspace{0.6cm} (k=1...6),
\end{eqnarray}
where $\hat{G}^{\tilde{\alpha}}_k$ and
$\hat{G}^{\tilde{\alpha}}_{5k}$ are the couplings in the
($\tilde{q}_L,\tilde{q}_R$) basis, and their explicit forms are
shown in Appendix A. The notations $a_k^\alpha$, $b_k^\alpha$
(k=1...6), and $a_{ik}^{\tilde{q}}$, $b_{ik}^{\tilde{q}}$
(k=1...4), and $l_{ik}^{\tilde{q}}$, $k_{ik}^{\tilde{q}}$ (k=1,2)
used in Eq.(\ref{lagrangian}) are defined also in Appendix A.

The tree-level amplitude of $\tilde{b}_i \rightarrow \tilde{t}_1
H^-$, as shown in Fig.1(a), is given by
\begin{eqnarray}
M^{(0)}_i=\frac{ig}{\sqrt{2}m_W}[R^{\tilde{b}}
\left(\begin{array}{cc} m_b^2\tan\beta +m_t^2\cot\beta
-m_W^2\sin2\beta
& m_t(A_t\cot\beta +\mu) \\
m_b(A_b\tan\beta +\mu) & 2m_tm_b/\sin2\beta
\end{array} \right) (R^{\tilde{t}})^T]_{i1},
\end{eqnarray}
and the decay width is
\begin{eqnarray}
\Gamma^{(0)}_i =\frac{|M^{(0)}_i|^2\lambda^{1/2}
(m_{\tilde{b}_i}^2, m_{\tilde{t}_1}^2, m_{H^-}^2)}{16\pi
m_{\tilde{b}_i}^3}
\end{eqnarray}
with $\lambda(x,y,z)=(x-y-z)^2-4yz$.

\section{Yukawa corrections}

The Feynman diagrams contributing to the Yukawa corrections to
$\tilde{b}_i \rightarrow \tilde{t}_1 H^-$ are shown in
Figs.1(b)--(f) and Fig.2. We carried out the calculation in the
t'Hooft-Feynman gauge and used the dimensional reduction, which
preserves supersymmetry, for regularization of the ultraviolet
divergences in the virtual loop corrections using the
on-mass-shell renormalization scheme\cite{on-mass}, in which the
fine-structure constant $\alpha_{ew}$ and physical masses are
chosen to be the renormalized parameters, and finite parts of the
counterterms are fixed by the renormalization conditions. The
coupling constant $g$ is related to the input parameters $e$,
$m_W$ and $m_Z$ via $g^2=e^2/s_w^2$ and $s_w^2=1-m_W^2/m_Z^2$. As
for the renormalization of the parameters in the Higgs sector and
the squark sector, it will be described below in detail.

The relevant renormalization constants are defined as
\begin{eqnarray}
&& m_{W0}^2=m_W^2 +\delta m_W^2, \hspace{0.5cm} m_{Z0}^2=m_Z^2
+\delta m_Z^2, \nonumber \\ && m_{q0}=m_q +\delta m_q,
\hspace{1.0cm} m_{\tilde{q}_i0}^2 =m_{\tilde{q}_i}^2 +\delta
m_{\tilde{q}_i}^2, \nonumber \\ && A_{q0}=A_q +\delta A_q,
\hspace{1.2cm} \mu_0 =\mu +\delta\mu, \nonumber \\ &&
\theta_{\tilde{q}0}=\theta_{\tilde{q}} +\delta\theta_{\tilde{q}},
\hspace{1.5cm} \tan\beta_0=(1 +\delta Z_\beta) \tan\beta,
\nonumber \\ && \tilde{q}_{i0}=(1 +\delta Z^{\tilde{q}}_i)^{1/2}
+\delta Z^{\tilde{q}}_{ij}\tilde{q}_j, \nonumber \\ && H_0^-=(1
+\delta Z_{H^-})^{1/2}H^- +\delta Z_{HG}G^-,  \nonumber \\ &&
G_0^-=(1 +\delta Z_{G^-})^{1/2}G^- +\delta Z_{GH}H^-
\end{eqnarray}
with $q=t,b$. Here we introduce the mixing of $H^-$ and
$G^-$\cite{GH}, instead of the mixing of $H^-$ and
$W^-$\cite{Mendez}
\begin{eqnarray}
&&  W^-_{\mu 0}=(1 +\delta Z_{W^-})^{1/2}W^-_{\mu}
+iZ_{WH}\partial_\mu H^-,  \nonumber    \\
&&  H_0^-=(1 +\delta Z_{H^-})^{1/2}H^-,
\end{eqnarray}
due to the reason: the former can successfully cancel the
divergences for the case we are considering, and the latter,
however, is just right for the renormalization of the parameters
$\beta$ and $\alpha$, and in the case that the particles
interacting with the charged Higgs boson are the on-shell fermions
(where a $\not{p}$ arising from $\partial_\mu H^-$ and the vertex
$W^-$--fermion--fermion is inserted between two on-shell fermions,
and turned into the fermion masses by Dirac equations, which is
therefore similar to the structure of the Yukawa coupling
$H^-$--fermion--fermion).

Taking into account the Yukawa corrections, the renormalized
amplitude for $\tilde{b}_i \rightarrow \tilde{t}_1 H^-$ is given
by
\begin{equation}
M^{ren}_i=M^{(0)}_i +\delta M^{(v)}_i +\delta M^{(c)}_i,
\end{equation}
where $\delta M^{(v)}_i$ and $\delta M^{(c)}_i$ are the vertex
corrections and  the counterterm, respectively.

The calculations of the vertex corrections from Fig.1(b)-1(f)
result in
\begin{eqnarray}
&&\delta M^{(v)}_i=
  \frac{i}{16\pi^2}\sum_{k=1}^6\sum_j(G_{5k}^{\tilde{b}})_{ij}
  (G_{k}^{\tilde{q}})_{j1}B_0(p_{\tilde{t}_1},m_{H_k},m_{\tilde{q}_j})
  \nonumber \\
  && \hspace{1.2cm}
  +\frac{i}{16\pi^2}\sum_{k=1}^6\sum_j(G_{k}^{\tilde{b}})_{ij}
  (G_{5k}^{\tilde{q}})_{j1}B_0(p_{\tilde{b}_i},m_{H_k},m_{\tilde{q}_j})
  \nonumber \\
  && \hspace{1.2cm}
  -\frac{i}{16\pi^2}\sum_{k=1}^{4}\sum_{j,l}(G_{k}^{\tilde{b}})_{ij}
  (G_{5}^{\tilde{b}})_{jl}(G_{k}^{\tilde{t}})_{l1}C_0(-p_{\tilde{b}_i},
  p_{H^-},m_{H_k},m_{\tilde{b}_j},m_{\tilde{t}_l}) \nonumber \\
  && \hspace{1.2cm}
  +\frac{ig^2}{8\pi^2}\sum_{k=1}^{4} \{[m_tm_ba_5^b +(m_{\tilde{b}_i}^2
  -p_{\tilde{b}_i}\cdot p_{H^-})b_5^b]m_{\tilde{\chi}_k^0}
  b_{ik}^{\tilde{b}}a_{1k}^{\tilde{t}\ast}C_0 \nonumber \\
  && \hspace{1.2cm}
  +b_{ik}^{\tilde{b}}[m_ba_5^bb_{1k}^{\tilde{t}\ast}(p_{\tilde{b}_i} -p_{H^-})
  +m_tb_5^bb_{1k}^{\tilde{t}\ast}p_{\tilde{b}_i}
  -m_{\tilde{\chi}_k^0}
  b_5^ba_{1k}^{\tilde{t}\ast}(2p_{\tilde{b}_i} -p_{H^-})]_{\mu}
  (-p_{\tilde{b}_i}^{\mu}C_{11} \nonumber \\
  && \hspace{1.2cm}
  +p_{H^-}^{\mu}C_{12})
  -(m_ba_5^bb_{1k}^{\tilde{t}\ast} +m_tb_5^bb_{1k}^{\tilde{t}\ast}
  -m_{\tilde{\chi}_k^0}b_5^ba_{1k}^{\tilde{t}\ast})b_{ik}^{\tilde{b}}
  (m_{\tilde{b}_i}^2C_{21} +m_{H^-}^2C_{22} \nonumber \\
  && \hspace{1.2cm}
  -2p_{\tilde{b}_i}\cdot p_{H^-}C_{23} +g_{\mu}^{\mu}C_{24})
  +(a\leftrightarrow b)\}(-p_{\tilde{b}_i},p_{H^-},m_{\tilde{\chi}_k^0},
  m_b,m_t) \nonumber \\
  && \hspace{1.2cm}
  -\frac{3i}{16\pi^2}\sum_{j,l}(h_b^2R_{l2}^{\tilde{b}}R_{i2}^{\tilde{b}}
  R_{j1}^{\tilde{t}}R_{11}^{\tilde{t}}+h_t^2R_{i1}^{\tilde{b}}R_{l1}^{\tilde{b}}
  R_{12}^{\tilde{t}}R_{j2}^{\tilde{t}})(G_5^{\tilde{b}})_{lj}
  B_0(p_{H^-},m_{\tilde{b}_l},m_{\tilde{t}_j}),
\end{eqnarray}
where $B_0$ and $C_{i(j)}$ are two- and three-point Feynman
integrals\cite{ABCD}, respectively, and $h_{t(b)}$ is the Yukawa
coupling defined by
\begin{equation}
h_t=\frac{gm_t}{\sqrt{2}m_W\sin\beta}, \hspace{1.0cm}
h_b=\frac{gm_b}{\sqrt{2}m_W\cos\beta}.
\end{equation}
In the first line of $\delta M^{(v)}_i$, $q=t$ for $k=1...4$ and
$q=b$ for $k=5,6$, respectively, while in the second line, $q=b$
for $k=1...4$ and $q=t$ for $k=5,6$, respectively.

The counterterm can be expressed as
\begin{eqnarray}
&& \delta M^{(c)}_i=i(\delta\theta_{\tilde b}+\delta
  Z_{21}^{\tilde b})(G_5^{\tilde b})_{3-i,1} +i(\delta\theta_{\tilde
  t}+\delta Z_{21}^{\tilde t})(G_5^{\tilde b})_{i2} +i[R^{\tilde
  b}(\delta \hat{G}_5^{\tilde b})(R^{\tilde t})^T]_{i1} \nonumber \\
&& \hspace{3.5cm}
  +\frac{i}{2}(\delta Z_i^{\tilde b}
  +\delta Z_1^{\tilde t} +\delta Z_{H^-})(G_5^{\tilde b})_{i1}
  +i\delta Z_{GH}(G_6^{\tilde b})_{i1}
\end{eqnarray}
with
\begin{eqnarray}
&& (\delta \hat{G}_5^{\tilde{b}})_{11}
  =\frac{g}{\sqrt{2}m_W}[(\frac{\delta g}{g} -\frac{1}{2}\frac{\delta
  m_W^2}{m_W^2})(m_b^2\tan\beta +m_t^2\cot\beta) +2m_b\tan\beta\delta
  m_b \nonumber \\
  && \hspace{1.5cm} +2m_t\cot\beta\delta m_t
  +m_b^2\delta \tan\beta +m_t^2\delta
  \cot\beta], \\
&& (\delta \hat{G}_5^{\tilde{b}})_{12}
  =\frac{g}{\sqrt{2}m_W}[(\frac{\delta g}{g} -\frac{1}{2}\frac{\delta m_W^2}{m_W^2}
  +\frac{\delta m_t}{m_t})m_t(A_t\cot\beta +\mu) +m_t(\delta
  A_t\cot\beta \nonumber \\
  && \hspace{1.5cm}
  +A_t\delta\cot\beta +\delta\mu)], \\
&& (\delta \hat{G}_5^{\tilde{b}})_{21}
  =\frac{g}{\sqrt{2}m_W}[(\frac{\delta g}{g} -\frac{1}{2}\frac{\delta m_W^2}{m_W^2}
  +\frac{\delta m_b}{m_b})m_b(A_b\tan\beta +\mu) +m_b(\delta
  A_b\tan\beta \nonumber \\
  && \hspace{1.5cm}
  +A_b\delta\tan\beta +\delta\mu)], \\
&& (\delta \hat{G}_5^{\tilde{b}})_{22}
  =\frac{2gm_bm_t}{\sqrt{2}m_W\sin{2\beta}}(\frac{\delta g}{g} -\frac{1}{2}\frac{\delta
  m_W^2}{m_W^2} +\frac{\delta m_b}{m_b} +\frac{\delta m_t}{m_t}
  -\cos{2\beta}\delta Z_\beta).
\end{eqnarray}
Here we consider only the corrections to the Yukawa couplings. The
explicit expressions of some renormalization constants calculated
from the self-energy diagrams in Fig.2 are given in Appendix B,
and other renormalization constants are fixed as follows.

For $\delta Z_{GH}$, using the approach discussed in the two-Higgs
doublet model (2HDM) in \cite{GH}, we derived below its expression
in the MSSM, where the version of the Higgs potential is different
from one of Ref. \cite{GH}. First, the one-loop renormalized
two-point function is given by
\begin{eqnarray}
i\Gamma_{GH}(p^2)=i(p^2 -m_{H^-}^2)\delta Z_{HG} +ip^2\delta
Z_{GH} -iT_{GH} +i\Sigma_{GH}(p^2),
\end{eqnarray}
where $T_{GH}$ is the tadpole function, which is given by
\begin{equation}
T_{GH}=\frac{g}{2m_W}[T_{H_2}\sin(\alpha -\beta)
+T_{H_1}\cos(\alpha -\beta)].
\end{equation}
Next from the on-shell renormalization condition, we obtained
\begin{equation}
\delta Z_{GH}=\frac{1}{m_{H^-}^2}[T_{GH} -\Sigma_{GH}(m_{H^-}^2)].
\end{equation}
The explicit expressions of $\Sigma_{GH}$ and the tadpole
counterterms $T_{H_k}$ $(k=1,2)$ are given in Appendix B.

For the renormalization of the parameter $\beta$, following the
analysis of Ref.\cite{Mendez}, we fixed the renormalization
constant by the requirement that the on-mass-shell $H^+ \bar{\tau}
\nu_\tau$ coupling remain the same form as in Eq.(3) of
Ref.\cite{Mendez} to all orders of perturbation theory. However,
with introducing the mixing of $H^-$ and $G^-$ instead of $H^-$
and $W^-$, the expression of $\delta Z_\beta$ is then changed to
\begin{eqnarray}
&& \delta Z_\beta=\frac{1}{2}\frac{\delta
  m_W^2}{m_W^2} -\frac{1}{2}\frac{\delta m_Z^2}{m_Z^2}
  +\frac{1}{2}\frac{\delta m_Z^2 -\delta m_W^2}{m_Z^2 -m_W^2}
  -\frac{1}{2}\delta Z_{H^+} +\cot\beta\delta Z_{GH}.
\end{eqnarray}

For the counterterm of squark mixing angle $\theta_{\tilde{q}}$,
using the same renormalized scheme as Ref.\cite{Yukawacorrection},
one has
\begin{equation}
\delta\theta_{\tilde{q}} =\frac{{\rm
Re}[\Sigma_{12}^{\tilde{q}}(m_{\tilde{q}_1}^2)
+\Sigma_{12}^{\tilde{q}}(m_{\tilde{q}_2}^2)]}{2(m_{\tilde{q}_1}^2
-m_{\tilde{q}_2}^2)}.
\end{equation}

For the renormalization of soft SUSY-breaking parameter $A_q$, we
fixed its counterterm by keeping the tree-level relation of $A_q$,
$m_{\tilde{q}_i}$ and $\theta_{\tilde{q}}$ \cite{Aq}, and get the
expression as following:
\begin{eqnarray}
&& \delta A_q=\frac{m_{\tilde{q}_1}^2-m_{\tilde{q}_2}^2}{2m_q}
  (2\cos{2\theta_{\tilde{q}}}\delta\theta_{\tilde{q}}
  -\sin{2\theta_{\tilde{q}}}\frac{\delta m_q}{m_q})
  +\frac{\sin{2\theta_{\tilde{q}}}}{2m_q}(\delta m_{\tilde{q}_1}^2
  -\delta m_{\tilde{q}_2}^2) \nonumber \\
  && \hspace{1.3cm}
  +\{\cot\beta, \tan\beta\}\delta \mu
  +\delta\{\cot\beta, \tan\beta\}\mu.
\end{eqnarray}

As for the parameter $\mu$, there are several
schemes\cite{full,mu,muonshell} to fix its counterterm, and here
we use the on-shell renormalization scheme in
Ref.\cite{muonshell}, which gives
\begin{equation}
\delta \mu= \sum_{k=1}^2[m_{\tilde{\chi}_k^+} (\delta U_{k2}V_{k2}
+U_{k2}\delta V_{k2}) +\delta m_{\tilde{\chi}_k^+}U_{k2}V_{k2}],
\end{equation}
where $(U,V)$ are the two $2\times 2$ matrices diagonalizing the
chargino mass matrix, and their counterterms $(\delta U,\delta V)$
are given by
\begin{eqnarray}
\delta U=\frac{1}{4}(\delta Z_R -\delta Z_R^T)U, \\
\delta V=\frac{1}{4}(\delta Z_L -\delta Z_L^T)V.
\end{eqnarray}
The mass shifts $\delta m_{\tilde{\chi}_k^+}$ and the off-diagonal
wave function renormalization constants $\delta Z_{R(L)}$ can be
written as
\begin{eqnarray}
&& \delta m_{\tilde{\chi}_k^+}=\frac{1}{2}{\rm
Re}[m_{\tilde{\chi}_k^+}(\Pi_{kk}^L(m_{\tilde{\chi}_k^+}^2)
+\Pi_{kk}^R(m_{\tilde{\chi}_k^+}^2))
+\Pi_{kk}^{S,L}(m_{\tilde{\chi}_k^+}^2)
+\Pi_{kk}^{S,R}(m_{\tilde{\chi}_k^+}^2)], \\
&& (\delta Z_R)_{ij}=\frac{2}{m_{\tilde{\chi}_i^+}^2
-m_{\tilde{\chi}_j^+}^2}{\rm Re}
[\Pi_{ij}^R(m_{\tilde{\chi}_j^+}^2)m_{\tilde{\chi}_j^+}^2
+\Pi_{ij}^L(m_{\tilde{\chi}_j^+}^2)m_{\tilde{\chi}_i^+}
m_{\tilde{\chi}_j^+} \nonumber \\ && \hspace{6.0cm}
+\Pi_{ij}^{S,R}(m_{\tilde{\chi}_j^+}^2) m_{\tilde{\chi}_i^+}
+\Pi_{ij}^{S,L}(m_{\tilde{\chi}_j^+}^2)m_{\tilde{\chi}_j^+}], \\
&& (\delta Z_L)_{ij}=(\delta Z_R)_{ij} \ \ (L\leftrightarrow R).
\end{eqnarray}
The explicit expressions of the chargino self-energy matrices
$\Pi^{L(R)}$ and $\Pi^{S,L(R)}$ are given in Appendix B.

Finally, the renormalized decay width is then given by
\begin{eqnarray}
\Gamma_i=\Gamma^{(0)}_i +\delta \Gamma^{(v)}_i +\delta
\Gamma^{(c)}_i
\end{eqnarray}
with
\begin{eqnarray}
\delta \Gamma^{(a)}_i =\frac{\lambda^{1/2}(m_{\tilde{b}_i}^2,
m_{\tilde{t}_1}^2, m_{H^-}^2)}{8\pi m_{\tilde{b}_i}^3} {\rm Re}
\{M^{(0)\ast}_i \delta M^{(a)}_i\} \ \ \ \ \ (a=v,c).
\end{eqnarray}

\section{Numerical results and conclusion}

In the following we present some numerical results for the Yukawa
corrections to sbottom decay into lighter stop plus charged Higgs
boson. In our numerical calculations the SM parameters were taken
to be $\alpha_{ew}(m_Z)=1/128.8$, $m_W=80.375$GeV,
$m_Z=91.1867$GeV\cite{SM} and $m_t=175.6$GeV. In order to improve
the convergence of the perturbation expansion, especially for
large $\tan\beta$, we take the running mass $m_b(Q)$ evaluated by
the next-to-leading order formula\cite{runningmb}
\begin{equation} \label{mbQ}
m_b(Q)=U_6(Q,m_t)U_5(m_t,m_b)m_b(m_b),
\end{equation}
where we have assumed that there are no other colored particles
with masses between scale $Q$ and $m_t$, and
$m_b(m_b)=4.25$GeV\cite{mb}. The evolution factor $U_f$ is
\begin{eqnarray}
U_f(Q_2,Q_1)=(\frac{\alpha_s(Q_2)}{\alpha_s(Q_1)})^{d^{(f)}}
[1+\frac{\alpha_s(Q_1)-\alpha_s(Q_2)}{4\pi}J^{(f)}], \nonumber \\
d^{(f)}=\frac{12}{33-2f}, \hspace{1.0cm}
J^{(f)}=-\frac{8982-504f+40f^2}{3(33-2f)^2},
\end{eqnarray}
where $\alpha_s(Q)$ is given by the solutions of the two-loop
renormalization group equations\cite{runningalphas}. When
$Q=400$GeV, the running mass $m_b(Q)\sim 2.5$GeV.  In addition, we
also improved the perturbation calculations by the following
replacement in the tree-level couplings\cite{runningmb}
\begin{equation}\label{replacement}
m_b(Q) \ \ \rightarrow \ \ \frac{m_b(Q)}{1+\Delta m_b(M_{SUSY})}
\end{equation}
where
\begin{eqnarray}\label{gluino}
&& \Delta m_b=\frac{2\alpha_s}{3\pi}M_{\tilde{g}}\mu\tan\beta
I(m_{\tilde{b}_1},m_{\tilde{b}_2},M_{\tilde{g}})
+\frac{h_t^2}{16\pi^2}\mu A_t\tan\beta
I(m_{\tilde{t}_1},m_{\tilde{t}_2},\mu) \nonumber \\
&& \hspace{1.0cm} -\frac{g^2}{16\pi^2}\mu M_2\tan\beta
[\cos^2\theta_{\tilde{t}}I(m_{\tilde{t}_1},M_2,\mu)
+\sin^2\theta_{\tilde{t}}I(m_{\tilde{t}_2},M_2,\mu) \nonumber \\
&& \hspace{4.0cm} +\frac{1}{2}\cos^2\theta_{\tilde{b}}
I(m_{\tilde{b}_1},M_2,\mu) +\frac{1}{2}\sin^2\theta_{\tilde{b}}
I(m_{\tilde{b}_2},M_2,\mu)]
\end{eqnarray}
with
\begin{eqnarray}
I(a,b,c)=\frac{1}{(a^2-b^2)(b^2-c^2)(a^2-c^2)}
(a^2b^2\log\frac{a^2}{b^2} +b^2c^2\log\frac{b^2}{c^2}
+c^2a^2\log\frac{c^2}{a^2}).
\end{eqnarray}

The two-loop leading-log relations\cite{Higgs} of the neutral
Higgs boson masses and mixing angles in the MSSM were used. For
$m_{H^+}$ the tree-level formula was used. Other MSSM parameters
were determined as follows:

(i) For the parameters $M_1$, $M_2$ and $\mu$ in the chargino and
neutralino matrix, we take $M_2$ and $\mu$ as the input
parameters, and then use the relation
$M_1=(5/3)(g'^2/g^2)M_2\simeq 0.5M_2$\cite{Haber,M1} to determine
$M_1$. The gluino mass $M_{\tilde{g}}$ in Eq.(\ref{gluino}) was
related to $M_2$ by
$M_{\tilde{g}}=(\alpha_s(M_{\tilde{g}})/\alpha_2)M_2$\cite{Hidaka}.

(ii) For the parameters $m^2_{\tilde{Q},\tilde{U},\tilde{D}}$ and
$A_{t,b}$ in squark mass matrices, we assumed $M_{\tilde
Q}=M_{\tilde U}=M_{\tilde D}$ and $A_t=A_b$ to simplify the
calculations.

Some typical numerical results of the tree-level decay widths and
the Yukawa corrections are given in Fig.3-9.

Fig.3 and Fig.4 show the $m_{\tilde{t}_1}$ dependence of the
results of $\tilde{b}_1$ and $\tilde{b}_2$ decays, respectively.
Here we take $m_{A^0}=150$GeV, $\mu=M_2=400$GeV and
$A_t=A_b=1$TeV. The leading terms of the tree level amplitude are
given by
\begin{eqnarray}\label{tree}
M^{(0)}_i\sim \frac{ig}{\sqrt{2}m_W}[m_t(A_t\cot\beta
+\mu)R^{\tilde{b}}_{i1}R^{\tilde{t}}_{12} +m_b(A_b\tan\beta
+\mu)R^{\tilde{b}}_{i2}R^{\tilde{t}}_{11}],
\end{eqnarray}
where $\cos\theta_{\tilde{b}}\sim$ (0.54, 0.67, 0.70) and
$\cos\theta_{\tilde{t}}\sim$ (-0.71, -0.71, -0.71) for
$\tan\beta=$ (4, 10, 30), respectively. In the case of $i=1$, two
terms in Eq.(\ref{tree}) have opposite signs, and their magnitudes
get close with the increasing $\tan\beta$ and thus cancel to large
extent for large $\tan\beta$. Therefore, the tree level decay
widths have the feature of $\Gamma_0(\tan\beta=4)
> \Gamma_0(\tan\beta=10) > \Gamma_0(\tan\beta=30)$ in the most of
parameter range, as shown in Fig.3(a). In the case of $i=2$, for
$\tan\beta=4,10$ and 30 two terms in Eq.(\ref{tree}) have the same
signs, so $\Gamma_0$ is larger than the one of the case of $i=1$.
From Fig.3(b) and Fig.4(b) one can see that the relative
corrections are sensitive to the value of $\tan\beta$. For
$\tan\beta=$ 30, the magnitudes of the corrections can exceed
$10\%$ when $m_{\tilde{t}_1}>160$GeV for $\tilde{b}_1$ decay and
$m_{\tilde{t}_1}>260$GeV for $\tilde{b}_2$ decay. Especially in
the case of $\tilde{b}_1$ decay it even can reach $40\%$, which is
due to the fact that the corresponding tree-level decay width
already becomes very small. There are the dips at
$m_{\tilde{t}_1}=$311GeV and 390GeV on the solid line in Fig.4(b),
which come from the singularities at the threshold points
$m_{\tilde{b}_2}=m_{\tilde{\chi}^+_1}+m_t$ and
$m_{\tilde{b}_1}=m_{\tilde{\chi}^+_1}+m_t$, respectively. However,
for $\tan\beta=$ 4 and 10, the corrections of two sbottom decays
are always small and range from $-5\%$ to $5\%$. In general, for
low $\tan\beta$ the top quark contribution is enhanced while for
high $\tan\beta$ the bottom quark contribution become large, and
for medium $\tan\beta$, there are not any enhanced effects from
Yukawa couplings. So the corrections for $\tan\beta=$ 4 are
generally larger than those for $\tan\beta$ =10, as shown in
Fig.3(b) and Fig.4(b).

Fig.5 (Fig.6) gives the tree-level decay width and the Yukawa
corrections as the functions of $m_{H^-}$ in the case of
$\tilde{b}_1$ decay ($\tilde{b}_2$ decay). We assumed
$m_{\tilde{t}_1}=170$GeV, $\mu=M_2=400$GeV and $A_t=A_b=1$TeV. The
features of the tree level decay widths in Fig.5(a) and Fig.6(a)
are similar to Fig.3(a) and Fig.4(a), respectively. From Fig.5(b)
and Fig.6(b) we can see that the relative corrections decrease or
increase the decay widths depending on $\tan\beta$. Fig.5(b) shows
that the corrections for $\tan\beta=4$ are always positive and
range between $6\%$ and $3\%$. For $\tan\beta=10$ the corrections
are negligiblely small. For high $\tan\beta(=30)$ the corrections
exceed $10\%$ when $m_{H^-}<180$GeV. There is a dip at
$m_{H^-}\sim 178$GeV on the solid curve due to the singularity of
the charged Higgs boson wave function renormalization constant at
the threshold point $m_{H^-}=m_t+m_b$. Fig.6(b) shows that the
corrections are a few percent for $\tan\beta=4,10$ and 30. There
are a dip and a peak on the solid curve, which arise from the
singularities at the threshold points
$m_{\tilde{b}_2}=m_{\tilde{b}_1}+m_{A^0}$ and $m_{H^-}=m_t+m_b$,
respectively.

In Fig.7 and Fig.8 we present the tree level decay widths and the
Yukawa corrections as the functions of $\mu$ in the case of
$\tilde{b}_1 \rightarrow \tilde {t}_1+ H^-$ and $\tilde{b}_2
\rightarrow \tilde {t}_1+ H^-$, respectively, assuming
$m_{\tilde{t}_1}=170$GeV, $M_2=400$GeV, $A_t=A_b=1$TeV and
$m_{A^0}=150$GeV. When $\mu$ takes some values, the tree level
decay widths are getting very small ($<10^{-3}$GeV), as shown in
Fig.7(a) and Fig.8(a), and the corrections near the above values
do not have a physical meaning. So we cut off the corrections,
since perturbation theory breaks down there. From Fig.7(a)
(Fig.8(a)) we can see that there is a high peak (a deep dip) for
$\tan\beta=30$ and $\mu\sim 30$GeV. This is due to the fact that
when $\tan\beta=30$ and $\mu\sim 30$GeV the second term in
Eq.(\ref{tree}) is enhanced (suppressed) greatly  for
$\sin\theta_{\tilde{b}}\sim 1$ ($\cos\theta_{\tilde{b}}\sim 0$) as
a result of the off-diagonal elements of $M_{\tilde{b}}^2$ in
Eq.(\ref{Mq2}) approaching to zero. Fig.7(b) and Fig.8(b) show
that the Yukawa corrections depend on $\mu$ strongly. Especially,
in the region of $\Gamma_0$ getting very small, the corrections
get the rapid variations between the positive and negative values
with the changes of the sign of the tree level amplitude. In
general, when the tree-level decay widths for $\tan\beta=$ 4 and
10 are not close to zero, the corrections are always small.
Comparing Fig.7(b) with Fig.8(b), we can find that the Yukawa
corrections for $\tan\beta=$ 30 are more significant for
$\tilde{b}_1$ decay than for $\tilde{b}_2$ decay.

Finally, in Fig.9 we show the Yukawa corrections as a function of
$\tan\beta$ in three ways of perturbative expansion: (i) the
strict on-shell scheme (the dashed line), where the bottom quark
pole mass 4.7GeV was used, (ii) the QCD-improved scheme (the
dotted line), in which only QCD running bottom quark mass $m_b(Q)$
in Eq.(\ref{mbQ}) was used, and (iii) the improved scheme (the
solid line), in which the replacement Eq.(\ref{replacement}) was
used. Here we assumed $m_{\tilde{t}_1}=170$GeV, $\mu=M_2=400$GeV,
$A_t=A_b=1$TeV and $m_{A^0}=150$GeV. One can see that the three
ways all give small corrections ($|\delta \Gamma/\Gamma_0|<5\%$)
for $\tan\beta$ ($<15$). However, the magnitude of the corrections
in the case (i) increases rapidly for $\tan\beta > 15$, and when
$\tan\beta>33$ the corrections will result in the physically
meaningless negative width. But the convergences in the cases (ii)
and (iii) are much better, and especially in the case (iii) the
magnitude of the corrections still is less than $40\%$ for high
values of $\tan\beta$ (=40).

In conclusion, we have calculated the Yukawa corrections to the
width of sbottom decay into lighter stop plus charged Higgs boson
in MSSM. These corrections depend on the masses of charged Higgs
boson and lighter stop, and the parameters $\tan\beta$ and $\mu$.
For favorable parameter values, the corrections decrease or
increase the decay widths significantly. Especially for high
values of $\tan\beta$(=30) the corrections exceed at least $10\%$
for both $\tilde b_1$ and $\tilde b_2$ decay. But for low values
of $\tan\beta$(=4,10) the corrections are small and the magnitudes
are less than $10\%$. The numerical calculations also show that
using the running bottom quark mass which includes the QCD effects
and resums all high order $\tan\beta$-enhanced effects, as given
in Ref.\cite{runningmb}, can improve much the convergence of the
perturbation expansion.

\section*{Acknowledgements}

We thank Ya Sheng Yang for giving help in numerical calculations.
This work was supported in part by the National Natural Science
Foundation of China, the Doctoral Program Foundation of Higher
Education of China, and a grant from the State Commission of
Science and Technology of China.

\section*{Appendix A}

The following couplings are given in order $O(h_t, h_b)$.

1. squark -- squark -- Higgs boson

(a) squark -- squark -- $h^0$
\begin{eqnarray}
\hat{G}^{\tilde{q}}_1= \left(\begin{array}{cc} -\sqrt{2}m_qh_q
\left\{\begin{array}{c}c_\alpha \\ -s_\alpha\end{array}\right\}
& -\frac{1}{\sqrt{2}}h_q(A_q \left\{\begin{array}{c}c_\alpha \\
-s_\alpha\end{array}\right\} +\mu \left \{\begin{array}{c}
s_\alpha \\ c_\alpha \end{array} \right\}) \\
-\frac{1}{\sqrt{2}}h_q(A_q \left\{\begin{array}{c} c_\alpha \\
-s_\alpha \end{array} \right\} +\mu \left \{\begin{array}{c}
s_\alpha \\ c_\alpha \end{array} \right\}) & -\sqrt{2}m_qh_q
\left\{\begin{array}{c} c_\alpha \\ -s_\alpha \end{array} \right\}
\end{array} \right)
\end{eqnarray}
for $\left\{\begin{array}{c} {\rm up} \\ {\rm down} \end{array}
\right\}$ type squarks, respectively. We use the abbreviations
$s_\alpha=\sin\alpha$, $c_\alpha=\cos\alpha$. $\alpha$ is the
mixing angle in the CP even neutral Higgs boson sector.

(b) squark -- squark -- $H^0$
\begin{eqnarray}
\hat{G}^{\tilde{q}}_2= \left(\begin{array}{cc} -\sqrt{2}m_qh_q
\left\{\begin{array}{c}s_\alpha \\ c_\alpha\end{array}\right\}
& -\frac{1}{\sqrt{2}}h_q(A_q \left\{\begin{array}{c} s_\alpha \\
c_\alpha\end{array}\right\} -\mu \left \{\begin{array}{c}
c_\alpha \\ s_\alpha \end{array} \right\}) \\
-\frac{1}{\sqrt{2}}h_q(A_q \left\{\begin{array}{c} s_\alpha \\
c_\alpha \end{array} \right\} -\mu \left \{\begin{array}{c}
c_\alpha \\ s_\alpha \end{array} \right\}) & -\sqrt{2}m_qh_q
\left\{\begin{array}{c} s_\alpha \\ c_\alpha \end{array} \right\}
\end{array} \right)
\end{eqnarray}

(c) squark -- squark -- $A^0$
\begin{eqnarray}
\hat{G}^{\tilde{q}}_3=i\frac{gm_q}{2m_W} \left(\begin{array}{cc} 0
& -A_q\left\{\begin{array}{c}\cot\beta \\ \tan\beta
\end{array}\right\} -\mu \\ A_q\left\{\begin{array}{c}\cot\beta
\\ \tan\beta \end{array}\right\} +\mu & 0 \end{array} \right)
\end{eqnarray}

(d) squark -- squark -- $G^0$
\begin{eqnarray}
\hat{G}^{\tilde{q}}_4=i\frac{gm_q}{2m_W} \left(\begin{array}{cc} 0
& -A_q +\mu\left\{\begin{array}{c}\cot\beta \\ \tan\beta
\end{array}\right\} \\ A_q -\mu\left\{\begin{array}{c}\cot\beta
\\ \tan\beta \end{array}\right\} & 0 \end{array} \right)
\end{eqnarray}

(e) squark -- squark -- $H^\pm$
\begin{eqnarray}
\hat{G}^{\tilde{b}}_5=(\hat{G}^{\tilde{t}}_5)^T =
\frac{g}{\sqrt{2}m_W}\left(\begin{array}{cc}
m_b^2\tan\beta +m_t^2\cot\beta & m_t(A_t\cot\beta +\mu) \\
m_b(A_b\tan\beta +\mu) & 2m_tm_b/\sin2\beta
\end{array} \right)
\end{eqnarray}

(f) squark -- squark -- $G^\pm$
\begin{eqnarray}
\hat{G}^{\tilde{b}}_6=(\hat{G}^{\tilde{t}}_6)^T =
\frac{g}{\sqrt{2}m_W}\left(\begin{array}{cc}
m_t^2 -m_b^2 & m_t(A_t -\mu\cot\beta) \\
m_b(\mu\tan\beta -A_b) & 0
\end{array} \right)
\end{eqnarray}

2. quark -- quark -- Higgs boson
\begin{eqnarray}
&& a_k^q=(\frac{1}{\sqrt{2}}h_q\left\{\begin{array}{c} -c_\alpha
\\ s_\alpha \end{array} \right\}, -\frac{1}{\sqrt{2}}h_q
\left\{\begin{array}{c} s_\alpha \\ c_\alpha \end{array} \right\},
-\frac{i}{\sqrt{2}}h_q\left\{\begin{array}{c} \cos\beta
\\ \sin\beta \end{array} \right\},
\frac{ig}{2m_W}\left\{\begin{array}{c} -m_t \\ m_b
\end{array}\right\}, \nonumber \\
&& \hspace{3.0cm} \left\{\begin{array}{c} h_b\sin\beta
\\ h_t\cos\beta \end{array} \right\},
\frac{g}{\sqrt{2}m_W}\left\{\begin{array}{c} -m_b \\m_t
\end{array}\right\})
\end{eqnarray}

\begin{eqnarray}
&& b_k^q=(\frac{1}{\sqrt{2}}h_q\left\{\begin{array}{c} -c_\alpha
\\ s_\alpha \end{array} \right\}, -\frac{1}{\sqrt{2}}h_q
\left\{\begin{array}{c} s_\alpha \\ c_\alpha \end{array} \right\},
-\frac{i}{\sqrt{2}}h_q\left\{\begin{array}{c} \cos\beta
\\ \sin\beta \end{array} \right\},
\frac{ig}{2m_W}\left\{\begin{array}{c} m_t \\ -m_b
\end{array}\right\}, \nonumber \\
&& \hspace{3.0cm} h_q \left\{\begin{array}{c} \cos\beta
\\ h_t\sin\beta \end{array} \right\},
\frac{g}{\sqrt{2}m_W}\left\{\begin{array}{c} m_t \\ -m_b
\end{array}\right\})
\end{eqnarray}

3. quark -- squark -- neutralino
\begin{eqnarray}
a_{ik}^{\tilde{q}}=-R_{i2}^{\tilde{q}}Y_q \left\{\begin{array}{c}
N_{k4} \\ N_{k3} \end{array} \right\}, \hspace{1.5cm}
b_{ik}^{\tilde{q}}=-R_{i1}^{\tilde{q}}Y_q \left\{\begin{array}{c}
N_{k4}^\ast \\ N_{k3}^\ast \end{array} \right\}
\end{eqnarray}
Here $N$ is the $4\times 4$ unitary matrix diagonalizing the
neutral gaugino-higgsino mass matrix \cite{Haber,M1}, and the
Yukawa factor $Y_q=h_q/g$.

4. quark -- squark -- chargino
\begin{eqnarray}
l_{ik}^{\tilde{q}}=R_{i2}^{\tilde{q}}Y_q \left\{\begin{array}{c}
V_{k2} \\ U_{k2} \end{array} \right\}, \hspace{1.5cm}
k_{ik}^{\tilde{q}}=R_{i1}^{\tilde{q}} \left\{\begin{array}{c}
Y_bU_{k2} \\ Y_tV_{k2} \end{array} \right\}.
\end{eqnarray}
Here $U$ and $V$ are the $2\times 2$ unitary matrices
diagonalizing the charged gaugino--higgsino mass matrix
\cite{Haber,M1}.

5. squark -- squark -- Higgs boson -- Higgs boson

(a) squark -- squark -- $H^-$ -- $H_k$ (k=1...4)
\begin{eqnarray}
\hat{G}_{5k}^{\tilde{b}}=(\hat{G}_{5k}^{\tilde{t}})^T
=\frac{g^2}{2\sqrt{2}m_W^2}\left(\begin{array}{cc} m_t^2S_k
+m_b^2T_k & 0 \\ 0 & 2m_tm_b/\sin2\beta V_k \end{array}\right)
\end{eqnarray}
with
\begin{eqnarray}
&& S_k=(\cos\alpha \cos\beta/\sin^2\beta, \ \ \  \sin\alpha
\cos\beta/\sin^2\beta,\ \ \ -i\cot^2\beta, \ \ \ i\cot\beta) \\ &&
T_k=(-\sin\alpha \sin\beta/\cos^2\beta, \ \ \  \cos\alpha
\sin\beta/\cos^2\beta, \ \ \ i\tan^2\beta, \ \ \  i\tan\beta) \\
&& V_k=(\sin(\beta -\alpha), \ \ \ \cos(\beta -\alpha), \ \ \  0,
\ \ \ i)
\end{eqnarray}

(b) squark -- squark -- $H^-$ -- $H^+$
\begin{eqnarray}
\hat{G}_{55}^{\tilde{q}}=\left(\begin{array}{cc}
-\left\{\begin{array}{c}h_b^2\sin^2\beta \\ h_t^2\cos^2\beta
\end{array}\right\} & 0 \\ 0 & -h_q^2\left\{\begin{array}{c}
\cos^2\beta \\ \sin^2\beta \end{array}\right\}
\end{array}\right)
\end{eqnarray}

(c) squark -- squark -- $H^-$ -- $G^+$
\begin{eqnarray}
\hat{G}_{56}^{\tilde{q}}=-\frac{g^2}{2m_W^2}\left(\begin{array}{cc}
\left\{\begin{array}{c}-m_b^2\tan\beta \\ m_t^2\cot\beta
\end{array} \right\} & 0 \\ 0 & m_q^2 \left\{\begin{array}{c}
\cot\beta \\ -\tan\beta \end{array} \right\}\end{array} \right)
\end{eqnarray}

\section*{Appendix B }

We define $q=t$ and $b$, $q'$ the $SU(2)_L$ partner of $q$, and
$q''=q$ for $k=1...4$ and $q''=q'$ for $k=5,6$. Then we have
\begin{eqnarray}
&& \frac{\delta m_W^2}{m_W^2}=\frac{g^2}{
  16\pi^2m_W^2}[m_b^2 +m_t^2 -A_0(m_t^2) -A_0(m_b^2) -m_t^2B_0 -(m_t^2
  -m_b^2)B_1] \nonumber \\
  && \hspace{1.3cm}
  (m_W^2,m_b,m_t), \nonumber \\
&& \frac{\delta m_Z^2}{m_Z^2}=\frac{3g^2}{8\pi^2m_W^2}
  \sum_{q=t,b}\{\frac{1}{3}[(I_{3L}^q -e_q\sin^2\theta_W)^2
  +e_q^2\sin^4\theta_W][2m_q^2 -2A_0(m_q^2) -m_q^2B_0]  \nonumber \\
  && \hspace{1.3cm}
  -2m_q^2e_q\sin^2\theta_W(I_{3L}^q -e_q\sin^2\theta_W)B_0\}
  (m_Z^2,m_q,m_q), \nonumber \\
&& \delta Z_{H^-}=\frac{3}{16\pi^2}\{2[(a_5^tb_5^b
  +b_5^ta_5^b)(m_{H^+}^2G_1 +B_1 +m_b^2G_0)
  +m_tm_b(a_5^ta_5^b+b_5^tb_5^b)G_0] \nonumber \\
  && \hspace{1.4cm}
  (m_{H^+}^2,m_b,m_t)
  -\sum_{j,l}(G_5^{\tilde{b}})_{jl}(G_5^{\tilde{t}})_{lj}
  G_0(m_{H^+}^2,m_{\tilde{b}_l},m_{\tilde{t}_j})\}, \nonumber \\
&& T_{H_k}=\frac{3}{16\pi^2}\sum_{q=t,b}\{2(a_k^q
  +b_k^q)m_qA_0(m_q^2) -\sum_j(G_k^{\tilde{q}})_{jj}A_0(m_{\tilde
  {q}_i}^2)\}, \nonumber \\
&& \Sigma_{GH}=-\frac{3}{16\pi^2}\{2[(a_5^tb_6^b+b_5^ta_6^b)
  (m_{H^+}^2B_1 +A_0(m_t^2) +m_b^2B_0) +m_tm_b(a_5^ta_6^b
  \nonumber \\
  && \hspace{1.3cm}
  +b_5^tb_6^b)B_0](m_{H^+}^2,m_b,m_t)
  -\sum_{j,l}(G_5^{\tilde{t}})_{jl}(G_6^{\tilde{b}})_{lj}B_0
  (m_{H^+}^2,m_{\tilde{b}_l},m_{\tilde{t}_j}) \nonumber \\
  && \hspace{1.3cm}
  +\sum_{q=t,b}\sum_j(G_{56}^{\tilde{q}})_{jj}A_0(m_{\tilde{q}_j}^2)\},
  \nonumber \\
&& \frac{\delta m_q}{m_q}=\frac{1}{16\pi^2}
  \{\sum_{k=1}^6[\frac{m_{q''}}{m_q} a_k^qa_k^{q''}B_0
  -\frac{1}{2}(a_k^qb_k^{q''}
  +b_k^qa_k^{q''})B_1](m_q^2,m_{q''},m_{H_k}) \nonumber \\
  && \hspace{1.3cm}
  +g^2\sum_{k=1}^4\sum_j[\frac{m_{\tilde{\chi}_k^0}}{m_q}
  a_{jk}^{\tilde{q}}b_{jk}^{\tilde{q}\ast}B_0
  -\frac{1}{2}(|a_{jk}^{\tilde{q}}|^2
  +|b_{jk}^{\tilde{q}}|^2)B_1](m_q^2,m_{\tilde{\chi}_k^0},m_{\tilde{q}_j})
  \nonumber \\
  && \hspace{1.3cm}
  +g^2\sum_{k=1}^2\sum_j[\frac{m_{\tilde{\chi}_k^+}}
  {m_q}l_{jk}^{\tilde{q}'}k_{jk}^{\tilde{q}^\prime\ast}B_0
  -\frac{1}{2}(|l_{jk}^{\tilde{q}'}|^2
  +|k_{jk}^{\tilde{q}^\prime}|^2)
  B_1](m_q^2,m_{\tilde{\chi}_k^+,m_{\tilde{q}_j^\prime}})\},
  \nonumber \\
&& \delta m_{\tilde{q}_i}^2=\frac{1}{16\pi^2}\{\sum_{k=1}^6\sum_j
  (G_k^{\tilde{q}})_{ij}(G_k^{\tilde{q}''})_{ji}B_0
  (m_{\tilde{q}_i}^2,m_{\tilde{q}_j''},m_{H_k})
  -2g^2\sum_{k=1}^4[(|a_{ik}^{\tilde{q}}|^2 +|b_{ik}^{\tilde{q}}|^2)
  \nonumber \\
  && \hspace{1.3cm}
  \times (m_{\tilde{q}_i}^2B_1
  +A_0(m_{\tilde{\chi}_k^0}^2) +m_q^2B_0)
  +2m_qm_{\tilde{\chi}_k^0}{\rm Re}(a_{ik}^{\tilde{q}}b_{ik}^{\tilde{q}\ast})
  B_0](m_{\tilde{q}_i}^2,m_q,m_{\tilde{\chi}_k^0}) \nonumber \\
  && \hspace{1.3cm}
  -2g^2\sum_{k=1}^2[(|l_{ik}^{\tilde{q}}|^2
  +|k_{ik}^{\tilde{q}}|^2)(m_{\tilde{q}_i'}^2B_1
  +A_0(m_{\tilde{\chi}_k^+}^2) +m_{q'}^2B_0) \nonumber \\
  && \hspace{1.3cm}
  +2m_{q'}m_{\tilde{\chi}_k^+}{\rm Re}(l_{ik}^{\tilde{q}}k_{ik}^{\tilde{q}\ast})
  B_0](m_{\tilde{q}_i}^2,m_{q'},m_{\tilde{\chi}_k^+})\}, \nonumber \\
&& \delta Z_{\tilde{q}_i}=\frac{1}{16\pi^2}\{-\sum_{k=1}^6\sum_j
  (G_k^{\tilde{q}})_{ij}(G_k^{\tilde{q}''})_{ji}G_0
  (m_{\tilde{q}_i}^2,m_{\tilde{q}_j''},m_{H_k})
  +2g^2\sum_{k=1}^4[(|a_{ik}^{\tilde{q}}|^2 +|b_{ik}^{\tilde{q}}|^2)
  \nonumber \\
  && \hspace{1.3cm}
  \times (B_1 +m_{\tilde{q}_i}^2G_1 +m_q^2G_0)
  +2m_qm_{\tilde{\chi}_k^0}{\rm Re}(a_{ik}^{\tilde{q}}b_{ik}^{\tilde{q}\ast})
  G_0](m_{\tilde{q}_i}^2,m_q,m_{\tilde{\chi}_k^0}) \nonumber \\
  && \hspace{1.3cm}
  +2g^2\sum_{k=1}^2[(|l_{ik}^{\tilde{q}}|^2
  +|k_{ik}^{\tilde{q}}|^2)(B_1 +m_{\tilde{q}_i'}^2G_1
  +m_{q'}^2G_0) \nonumber \\
  && \hspace{1.3cm}
  +2m_{q'}m_{\tilde{\chi}_k^+}{\rm Re}(l_{ik}^{\tilde{q}}k_{ik}^{\tilde{q}\ast})
  G_0](m_{\tilde{q}_i}^2,m_{q'},m_{\tilde{\chi}_k^+})\},
  \nonumber \\
&& \Sigma_{12}^{\tilde{q}}(p^2)=\frac{1}{16\pi^2}
  \{\sum_{k=1}^6\sum_j
  (G_k^{\tilde{q}})_{1j}(G_k^{\tilde{q}''})_{j2}B_0
  (p^2,m_{\tilde{q}_j''},m_{H_k})
  -2g^2\sum_{k=1}^4[(a_{1k}^{\tilde{q}}a_{2k}^{\tilde{q}\ast}
  +b_{1k}^{\tilde{q}}b_{2k}^{\tilde{q}\ast})  \nonumber \\
  && \hspace{1.3cm}
  \times (p^2B_1 +A_0(m_{\tilde{\chi}_k^0}^2) +m_q^2B_0)
  +m_qm_{\tilde{\chi}_k^0}(a_{1k}^{\tilde{q}}b_{2k}^{\tilde{q}\ast}
  +a_{2k}^{\tilde{q}\ast}b_{1k}^{\tilde{q}})
  B_0](p^2,m_q,m_{\tilde{\chi}_k^0}) \nonumber \\
  && \hspace{1.3cm}
  -2g^2\sum_{k=1}^2[(l_{1k}^{\tilde{q}}l_{2k}^{\tilde{q}\ast}
  +k_{1k}^{\tilde{q}}k_{2k}^{\tilde{q}\ast})(p^2B_1
  +A_0(m_{\tilde{\chi}_k^+}^2) +m_{q'}^2B_0) \nonumber \\
  && \hspace{1.3cm}
  +m_{q'}m_{\tilde{\chi}_k^+}(l_{1k}^{\tilde{q}}k_{2k}^{\tilde{q}\ast}
  +l_{2k}^{\tilde{q}\ast}k_{1k}^{\tilde{q}})
  B_0](p^2,m_{q'},m_{\tilde{\chi}_k^+})\}, \nonumber \\
&& \delta \theta_{\tilde{q}} +\delta Z_{21}^{\tilde{q}}
  =\frac{1}{2(m_{\tilde{q}_1}^2
  -m_{\tilde{q}_2}^2)}[\Sigma_{12}^{\tilde{q}}(m_{\tilde{q}_2}^2)
  -\Sigma_{12}^{\tilde{q}}(m_{\tilde{q}_1}^2)], \nonumber \\
&& \Pi_{ij}^L(p^2)=-\frac{3}{16\pi^2}\sum_{k=1}^2
[l_{ki}^{\tilde{t}} l_{kj}^{\tilde{t}}B_1(p^2,m_b,m_{\tilde{t}_k})
+k_{ki}^{\tilde{b}}
k_{kj}^{\tilde{b}}B_1(p^2,m_t,m_{\tilde{b}_k})], \nonumber \\
&& \Pi_{ij}^R(p^2)=-\frac{3}{16\pi^2}\sum_{k=1}^2
[k_{ki}^{\tilde{t}} k_{kj}^{\tilde{t}}B_1(p^2,m_b,m_{\tilde{t}_k})
+l_{ki}^{\tilde{b}}
l_{kj}^{\tilde{b}}B_1(p^2,m_t,m_{\tilde{b}_k})], \nonumber \\
&& \Pi_{ij}^{S,L}(p^2)=\frac{3}{16\pi^2}\sum_{k=1}^2
[m_bk_{ki}^{\tilde{t}}
l_{kj}^{\tilde{t}}B_0(p^2,m_b,m_{\tilde{t}_k})
+m_tl_{ki}^{\tilde{b}}
k_{kj}^{\tilde{b}}B_0(p^2,m_t,m_{\tilde{b}_k})], \nonumber \\
&& \Pi_{ij}^{S,R}(p^2)=\frac{3}{16\pi^2}\sum_{k=1}^2
[m_bl_{ki}^{\tilde{t}}
k_{kj}^{\tilde{t}}B_0(p^2,m_b,m_{\tilde{t}_k})
+m_tk_{ki}^{\tilde{b}}
l_{kj}^{\tilde{b}}B_0(p^2,m_t,m_{\tilde{b}_k})]. \nonumber
\end{eqnarray}
Here $A_0$ and $B_1$ are one- and two-point Feynman
integrals\cite{ABCD}, respectively, and $G_i=\partial B_i/\partial
p^2$.


\newpage
\begin{figure}
\begin{center}
\begin{picture}(110,110)(0,0)
\DashLine(20,60)(55,60){3} \DashLine(55,60)(85,90){3}
\DashLine(55,60)(85,30){3} \Vertex(55,60){1}
\Text(15,66)[]{$\tilde{b}_i$} \Text(93,91)[]{$\tilde{t}_1$}
\Text(93,29)[]{$H^-$} \Text(55,10)[]{$(a)$}
\end{picture}
\hspace{0.8cm}
\begin{picture}(110,110)(0,0)
\DashLine(15,60)(45,60){3} \DashLine(45,60)(70,77.5){3}
\DashLine(70,77.5)(70,42.5){3} \DashLine(70,42.5)(45,60){3}
\DashLine(70,77.5)(95,95){3} \DashLine(70,42.5)(95,25){3}
\Vertex(45,60){1} \Vertex(70,77.5){1} \Vertex(70,42.5){1}
\Text(9,66)[]{$\tilde{b}_i$} \Text(55,80)[]{$H_k$}
\Text(55,42)[]{$\tilde{b}_j$} \Text(78,60)[]{$\tilde{b}_l$}
\Text(103,96)[]{$\tilde{t}_1$} \Text(103,24)[]{$H^-$}
\Text(55,10)[]{$(b)$}
\end{picture}
\hspace{0.8cm}
\begin{picture}(110,110)(0,0)
\DashLine(15,60)(45,60){3} \Line(45,60)(70,77.5)
\ArrowLine(70,42.5)(70,77.5) \ArrowLine(45,60)(70,42.5)
\DashLine(70,77.5)(95,95){3} \DashLine(70,42.5)(95,25){3}
\Vertex(45,60){1} \Vertex(70,77.5){1} \Vertex(70,42.5){1}
\Text(9,66)[]{$\tilde{b}_i$} \Text(55,80)[]{$\tilde{\chi}_k^0$}
\Text(55,45)[]{$b$} \Text(78,60)[]{$t$}
\Text(103,96)[]{$\tilde{t}_1$} \Text(103,24)[]{$H^-$}
\Text(55,10)[]{$(c)$}
\end{picture}
\end{center}

\begin{center}
\begin{picture}(110,110)(0,0)
\DashLine(20,60)(55,60){3} \DashCArc(63.5,68.5)(12,0,360){3}
\DashLine(72,77)(85,90){3} \DashLine(55,60)(85,30){3}
\Vertex(55,60){1} \Vertex(72,77){1} \Text(15,66)[]{$\tilde{b}_i$}
\Text(93,91)[]{$\tilde{t}_1$} \Text(93,29)[]{$H^-$}
\Text(43,80)[]{$H_k$} \Text(83,63)[]{$\tilde{q}_j$}
\Text(55,10)[]{$(d)$}
\end{picture}
\hspace{1.0cm}
\begin{picture}(110,110)(0,0)
\DashLine(12,60)(31,60){3} \DashCArc(43,60)(12,0,360){3}
\DashLine(55,60)(85,90){3} \DashLine(55,60)(85,30){3}
\Vertex(55,60){1} \Vertex(31,60){1} \Text(7,66)[]{$\tilde{b}_i$}
\Text(93,91)[]{$\tilde{t}_1$} \Text(93,29)[]{$H^-$}
\Text(43,80)[]{$H_k$} \Text(43,40)[]{$\tilde{q}_j$}
\Text(55,10)[]{$(e)$}
\end{picture}
\hspace{0.8cm}
\begin{picture}(110,110)(0,0)
\DashLine(20,60)(55,60){3} \DashLine(55,60)(85,90){3}
\DashCArc(63.5,51.5)(12,0,360){3} \DashLine(72,43)(85,30){3}
\Vertex(55,60){1} \Vertex(72,43){1} \Text(15,66)[]{$\tilde{b}_i$}
\Text(93,91)[]{$\tilde{t}_1$} \Text(93,29)[]{$H^-$}
\Text(45,42)[]{$\tilde{b}_l$} \Text(83,63)[]{$\tilde{t}_j$}
\Text(55,10)[]{$(f)$}
\end{picture}
\end{center}
\caption{Feynman diagrams contributing to supersymmetric
electroweak corrections to $\tilde{b}_i\rightarrow \tilde{t}_1
H^-$: $(a)$ is tree level diagram; $(b)-(f)$ are one-loop vertex
corrections. In diagram $(b)$ the subscript k of $H_k$ can take
from 1 to 4. In diagram $(d)$ $q=t$ for $k=1...4$ and $q=b$ for
$k=5,6$. In diagram $(e)$ $q=b$ for $k=1...4$ and $q=t$ for
$k=5,6$.} \label{vertex}
\end{figure}
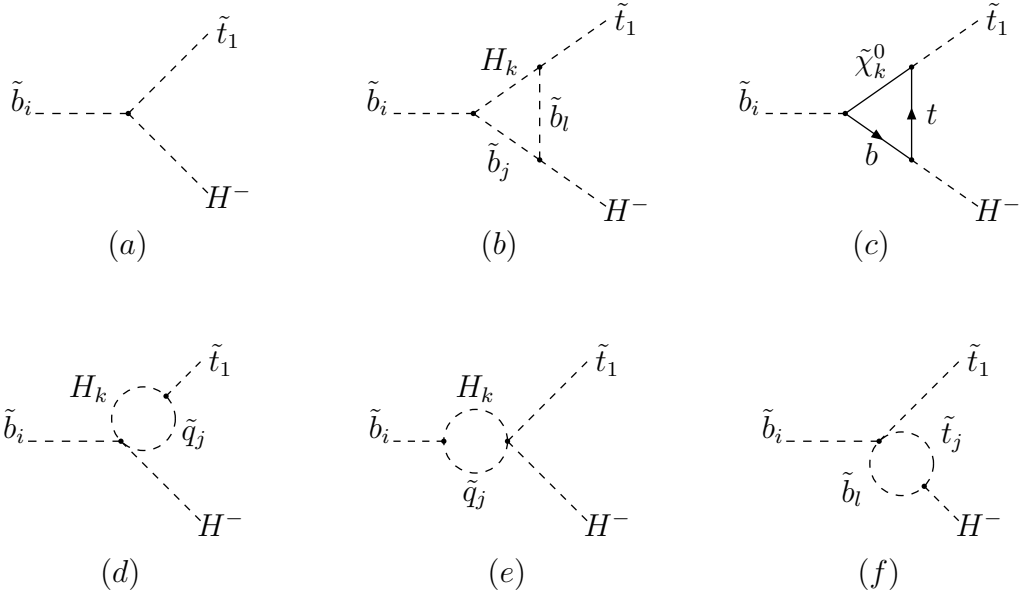
\newpage
\begin{figure}
\begin{center}
\begin{picture}(100,100)(0,0)
\Photon(10,50)(34,50){2}{3} \ArrowArc(50,50)(16,0,180)
\ArrowArc(50,50)(16,180,360) \Photon(66,50)(90,50){2}{3}
\Vertex(34,50){1} \Vertex(66,50){1} \Text(7,40)[]{$Z^0$}
\Text(50,76)[]{$t(b)$} \Text(50,24)[]{$t(b)$}
\Text(93,40)[]{$Z^0$} \Text(50,5)[]{$(a)$}
\end{picture}
\hspace{1.0cm}
\begin{picture}(100,100)(0,0)
\Photon(10,50)(34,50){2}{3} \ArrowArc(50,50)(16,0,180)
\ArrowArc(50,50)(16,180,360) \Photon(66,50)(90,50){2}{3}
\Vertex(34,50){1} \Vertex(66,50){1} \Text(10,40)[]{$W^-$}
\Text(50,76)[]{$t$} \Text(50,24)[]{$b$} \Text(93,40)[]{$W^-$}
\Text(50,5)[]{$(b)$}
\end{picture}
\hspace{1.0cm}
\begin{picture}(100,100)(0,0)
\DashLine(10,50)(34,50){3} \ArrowArc(50,50)(16,0,180)
\ArrowArc(50,50)(16,180,360) \DashLine(66,50)(90,50){3}
\Vertex(34,50){1} \Vertex(66,50){1} \Text(10,40)[]{$H^-$}
\Text(50,76)[]{$t$} \Text(50,24)[]{$b$} \Text(93,40)[]{$H^-(G^-)$}
\Text(50,5)[]{$(c)$}
\end{picture}
\end{center}

\begin{center}
\begin{picture}(100,100)(0,0)
\DashLine(10,50)(34,50){3} \DashCArc(50,50)(16,0,360){3}
\DashLine(66,50)(90,50){3} \Vertex(34,50){1} \Vertex(66,50){1}
\Text(10,40)[]{$H^-$} \Text(50,76)[]{$\tilde{t}_i$}
\Text(50,24)[]{$\tilde{b}_j$} \Text(93,40)[]{$H^-(G^-)$}
\Text(50,5)[]{$(d)$}
\end{picture}
\hspace{1.0cm}
\begin{picture}(100,100)(0,0)
\DashLine(10,40)(90,40){3} \DashCArc(50,54)(14,0,360){3}
\Vertex(50,40){1} \Text(10,30)[]{$H^-$}
\Text(50,80)[]{$\tilde{t}_i,\tilde{b}_i$}
\Text(93,30)[]{$H^-(G^-)$} \Text(50,5)[]{$(e)$}
\end{picture}
\hspace{1.0cm}
\begin{picture}(100,100)(0,0)
\DashLine(10,50)(34,50){3} \DashCArc(50,50)(16,0,360){3}
\DashLine(66,50)(90,50){3} \Vertex(34,50){1} \Vertex(66,50){1}
\Text(10,40)[]{$\tilde{t}_i(\tilde{b}_i)$} \Text(50,76)[]{$H_k$}
\Text(50,24)[]{$\tilde{q}_l$}
\Text(93,40)[]{$\tilde{t}_j(\tilde{b}_j)$} \Text(50,5)[]{$(f)$}
\end{picture}
\end{center}

\begin{center}
\begin{picture}(100,100)(0,0)
\DashLine(10,50)(34,50){3} \CArc(50,50)(16,0,360)
\DashLine(66,50)(90,50){3} \Vertex(34,50){1} \Vertex(66,50){1}
\Text(10,40)[]{$\tilde{t}_i(\tilde{b}_i)$}
\Text(50,76)[]{$\tilde{\chi}^0_k$} \Text(50,24)[]{$t(b)$}
\Text(93,40)[]{$\tilde{t}_j(\tilde{b}_j)$} \Text(50,5)[]{$(g)$}
\end{picture}
\hspace{1.0cm}
\begin{picture}(100,100)(0,0)
\DashLine(10,50)(34,50){3} \CArc(50,50)(16,0,360)
\DashLine(66,50)(90,50){3} \Vertex(34,50){1} \Vertex(66,50){1}
\Text(10,40)[]{$\tilde{t}_i(\tilde{b}_i)$}
\Text(50,76)[]{$\tilde{\chi}^+_k$} \Text(50,24)[]{$b(t)$}
\Text(93,40)[]{$\tilde{t}_j(\tilde{b}_j)$} \Text(50,5)[]{$(h)$}
\end{picture}
\hspace{1.0cm}
\begin{picture}(100,100)(0,0)
\DashLine(10,40)(90,40){3} \DashCArc(50,54)(14,0,360){3}
\Vertex(50,40){1} \Text(10,30)[]{$\tilde{t}_i(\tilde{b}_i)$}
\Text(50,80)[]{$H_k,\tilde{t}_l,\tilde{b}_l$}
\Text(93,30)[]{$\tilde{t}_j(\tilde{b}_j)$} \Text(50,5)[]{$(i)$}
\end{picture}
\end{center}

\begin{center}
\begin{picture}(100,100)(0,0)
\ArrowLine(10,50)(34,50) \ArrowLine(34,50)(66,50)
\ArrowLine(66,50)(90,50) \DashCArc(50,50)(16,0,180){3}
\Vertex(34,50){1} \Vertex(66,50){1} \Text(10,40)[]{$t(b)$}
\Text(50,40)[]{$q$} \Text(50,75)[]{$H_k$} \Text(93,40)[]{$t(b)$}
\Text(50,5)[]{$(j)$}
\end{picture}
\hspace{1.0cm}
\begin{picture}(100,100)(0,0)
\ArrowLine(10,50)(34,50) \Line(34,50)(66,50)
\ArrowLine(66,50)(90,50) \DashCArc(50,50)(16,0,180){3}
\Vertex(34,50){1} \Vertex(66,50){1} \Text(10,40)[]{$t(b)$}
\Text(50,40)[]{$\tilde{\chi}^0_k$}
\Text(50,75)[]{$\tilde{t}_i(\tilde{b}_i)$} \Text(93,40)[]{$t(b)$}
\Text(50,5)[]{$(k)$}
\end{picture}
\hspace{1.0cm}
\begin{picture}(100,100)(0,0)
\ArrowLine(10,50)(34,50) \Line(34,50)(66,50)
\ArrowLine(66,50)(90,50) \DashCArc(50,50)(16,0,180){3}
\Vertex(34,50){1} \Vertex(66,50){1} \Text(10,40)[]{$t(b)$}
\Text(50,40)[]{$\tilde{\chi}^+_k$}
\Text(50,75)[]{$\tilde{b}_i(\tilde{t}_i)$} \Text(93,40)[]{$t(b)$}
\Text(50,5)[]{$(l)$}
\end{picture}
\end{center}

\begin{center}
\begin{picture}(100,100)(0,0)
\ArrowLine(10,50)(34,50) \Line(34,50)(66,50)
\ArrowLine(66,50)(90,50) \DashCArc(50,50)(16,0,180){3}
\Vertex(34,50){1} \Vertex(66,50){1}
\Text(10,40)[]{$\tilde{\chi}^+_i$} \Text(50,40)[]{$t(b)$}
\Text(50,75)[]{$\tilde{b}_k(\tilde{t}_k)$}
\Text(93,40)[]{$\tilde{\chi}^+_j$} \Text(50,5)[]{$(m)$}
\end{picture}
\hspace{1.0cm}
\begin{picture}(100,100)(0,0)
\DashLine(50,55)(50,30){3} \ArrowArc(50,70)(15,0,180)
\CArc(50,70)(15,180,360) \Vertex(50,55){1}
\Text(75,40)[]{$H^0,h^0$} \Text(50,95)[]{$t,b$}
\Text(50,5)[]{$(n)$}
\end{picture}
\hspace{1.0cm}
\begin{picture}(100,100)(0,0)
\DashLine(50,55)(50,30){3} \DashCArc(50,70)(15,0,360){3}
\Vertex(50,55){1} \Text(75,40)[]{$H^0,h^0$}
\Text(50,95)[]{$\tilde{t}_i,\tilde{b}_i$} \Text(50,5)[]{$(o)$}
\end{picture}
\end{center}
\caption{Feynman diagrams contributing to renormalization
constants. In diagram $(i)$ $q=t(b)$ for $k=1...4$ and $q=b(t)$
for $k=5,6$.} \label{self}
\end{figure}
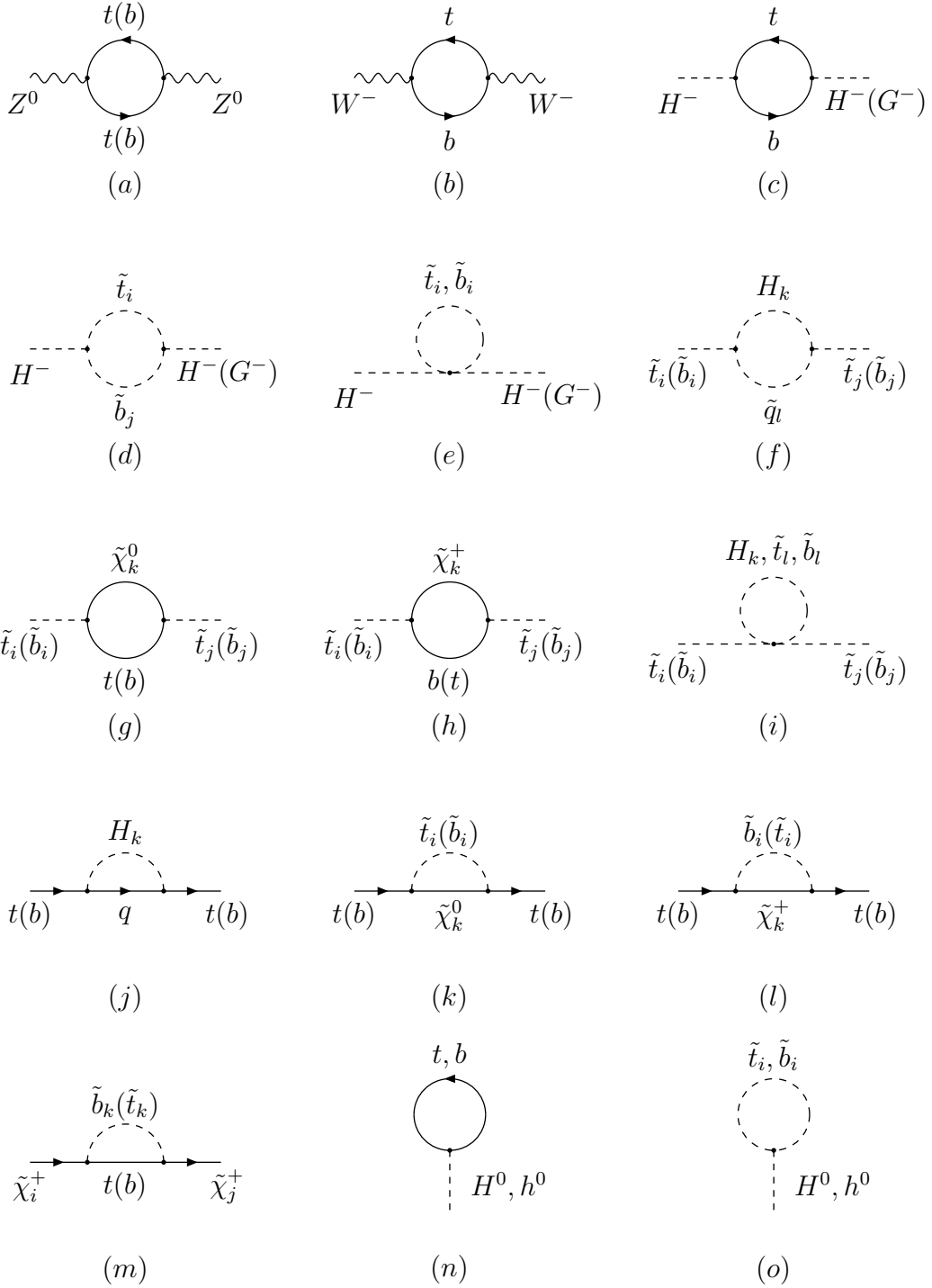
\newpage
\begin{figure}
\centerline{\epsfig{file=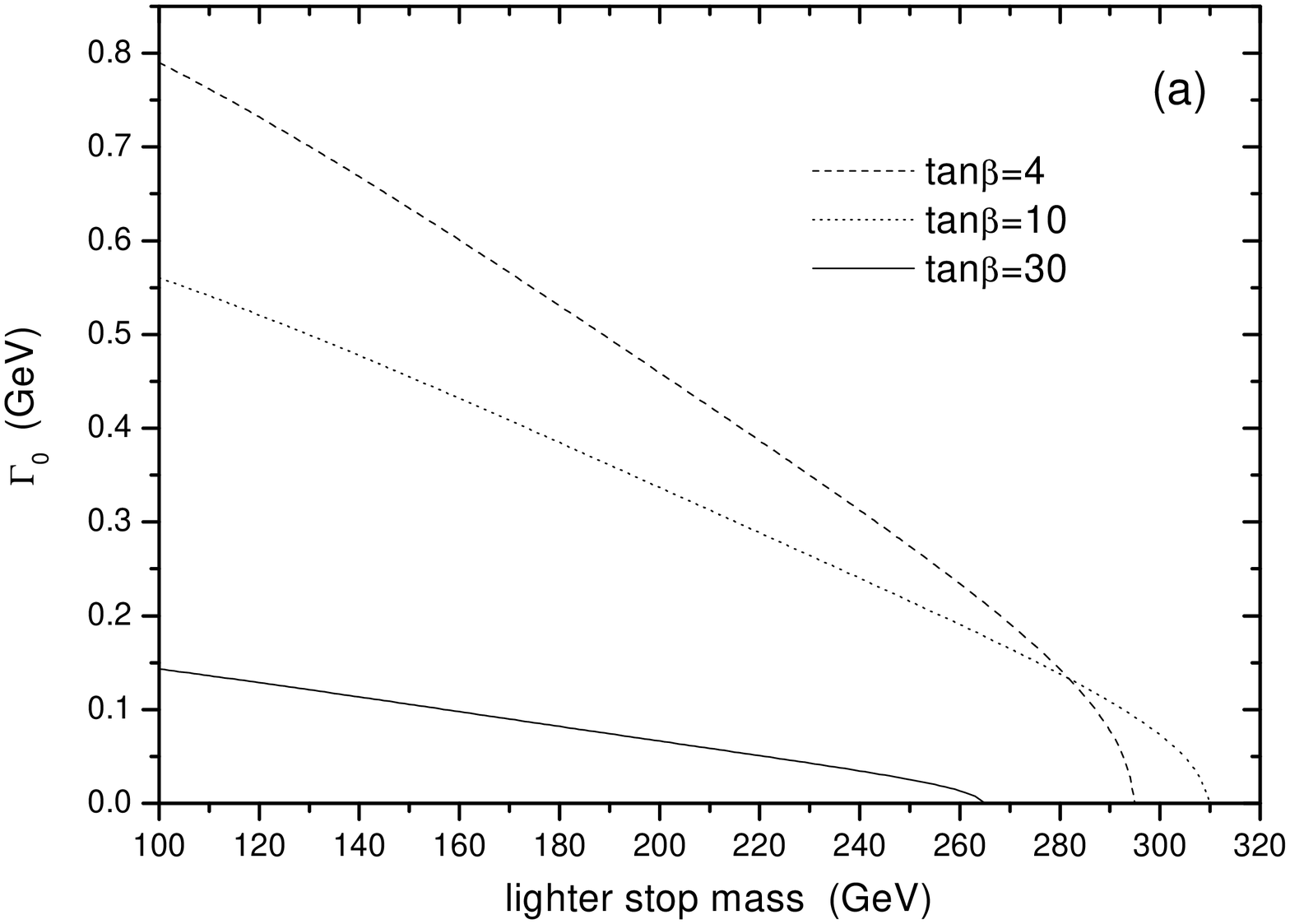, width=400pt}}
\centerline{\epsfig{file=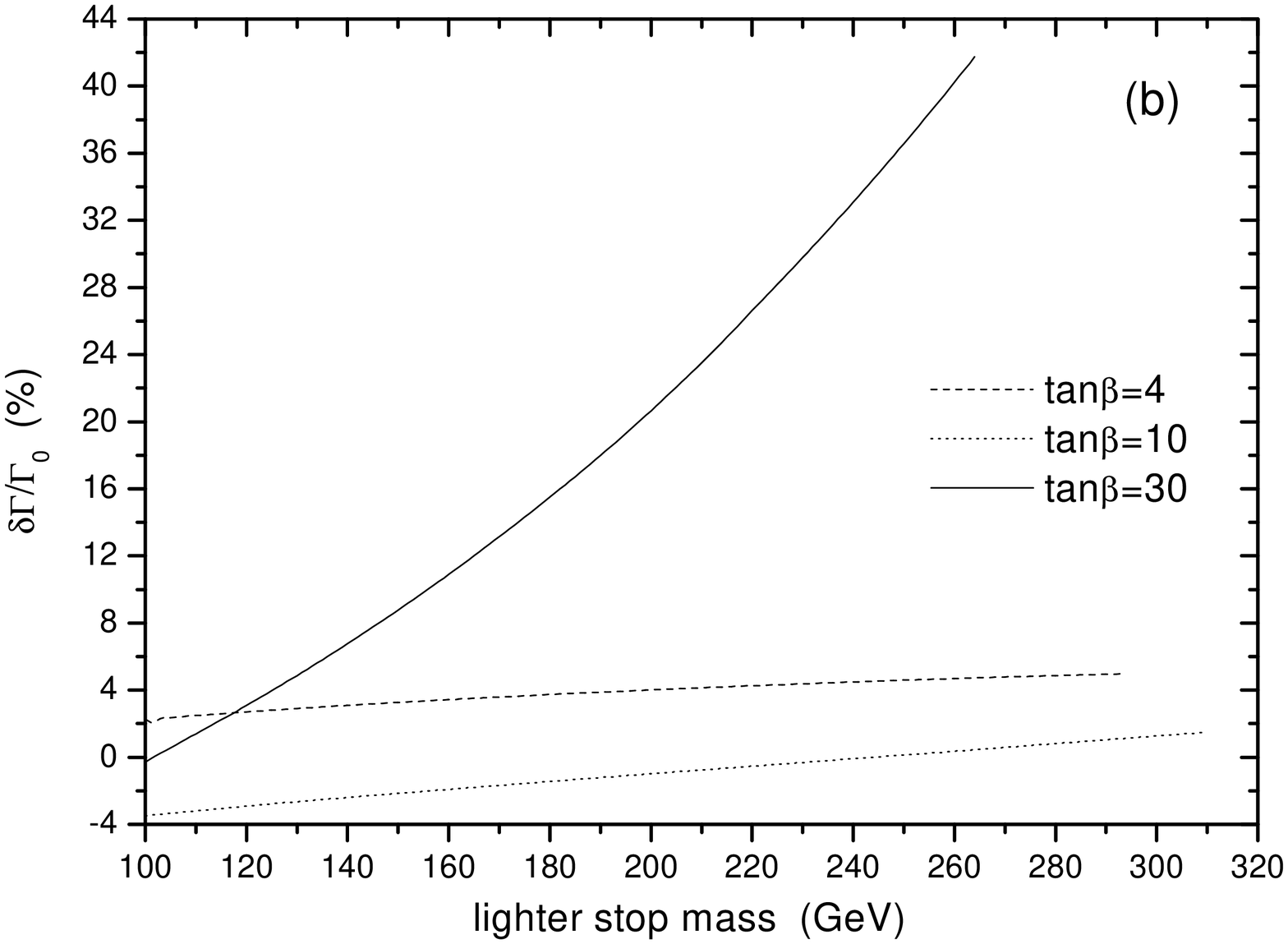, width=400pt}} \caption{The
tree-level decay width (figure $(a)$) of $\tilde{b}_1 \rightarrow
\tilde{t}_1H^-$ and its Yukawa corrections (figure $(b)$) as
functions of $m_{\tilde{t}_1}$ for $\tan\beta=4$, $10$ and $30$,
respectively, assuming $m_{A^0}=150$GeV, $\mu=M_2=400$GeV,
$A_t=A_b=1$TeV and $M_{\tilde Q}=M_{\tilde U}=M_{\tilde D}$.}
\label{mst1}
\end{figure}
\begin{figure}
\centerline{\epsfig{file=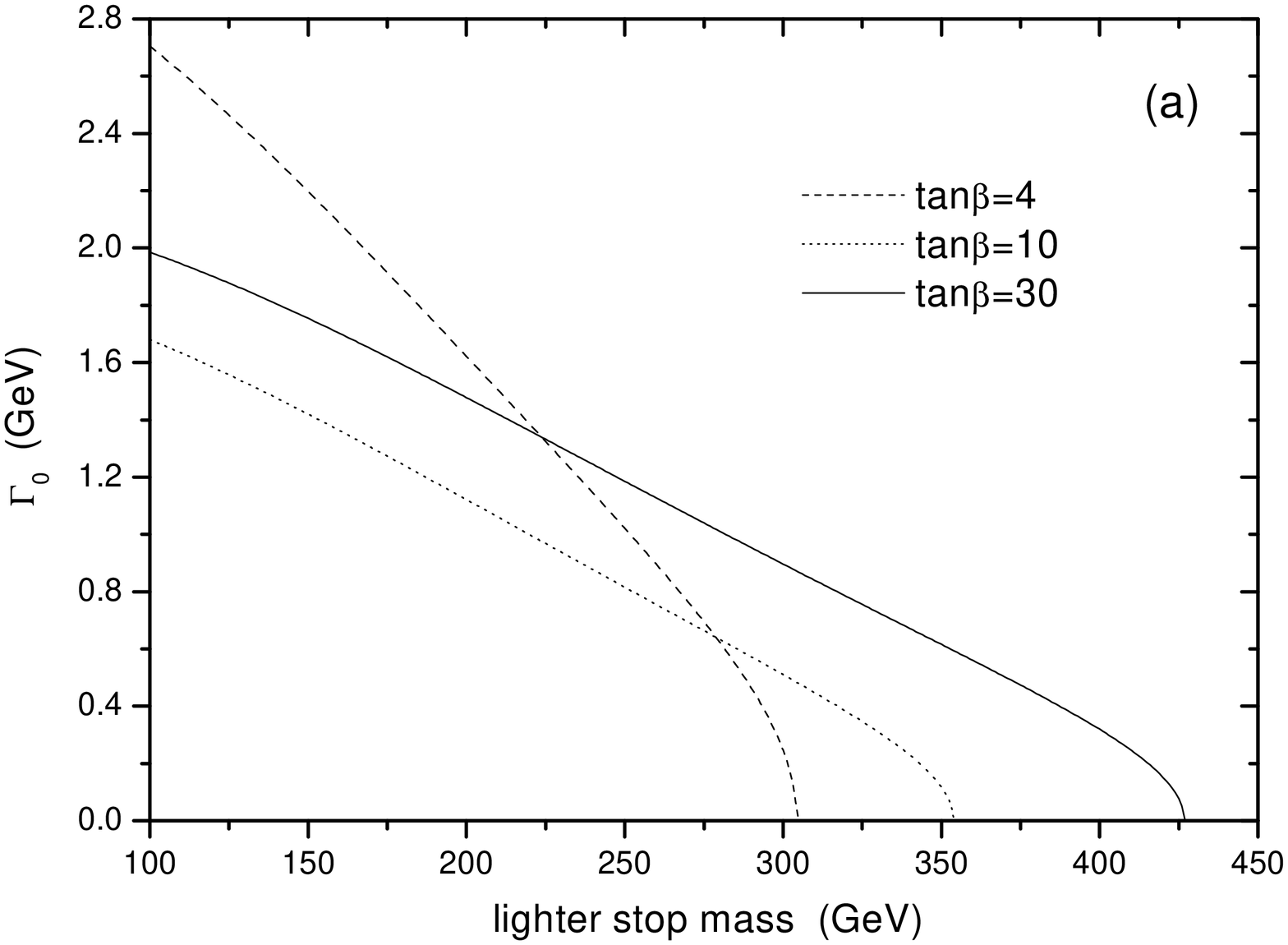, width=400pt}}
\centerline{\epsfig{file=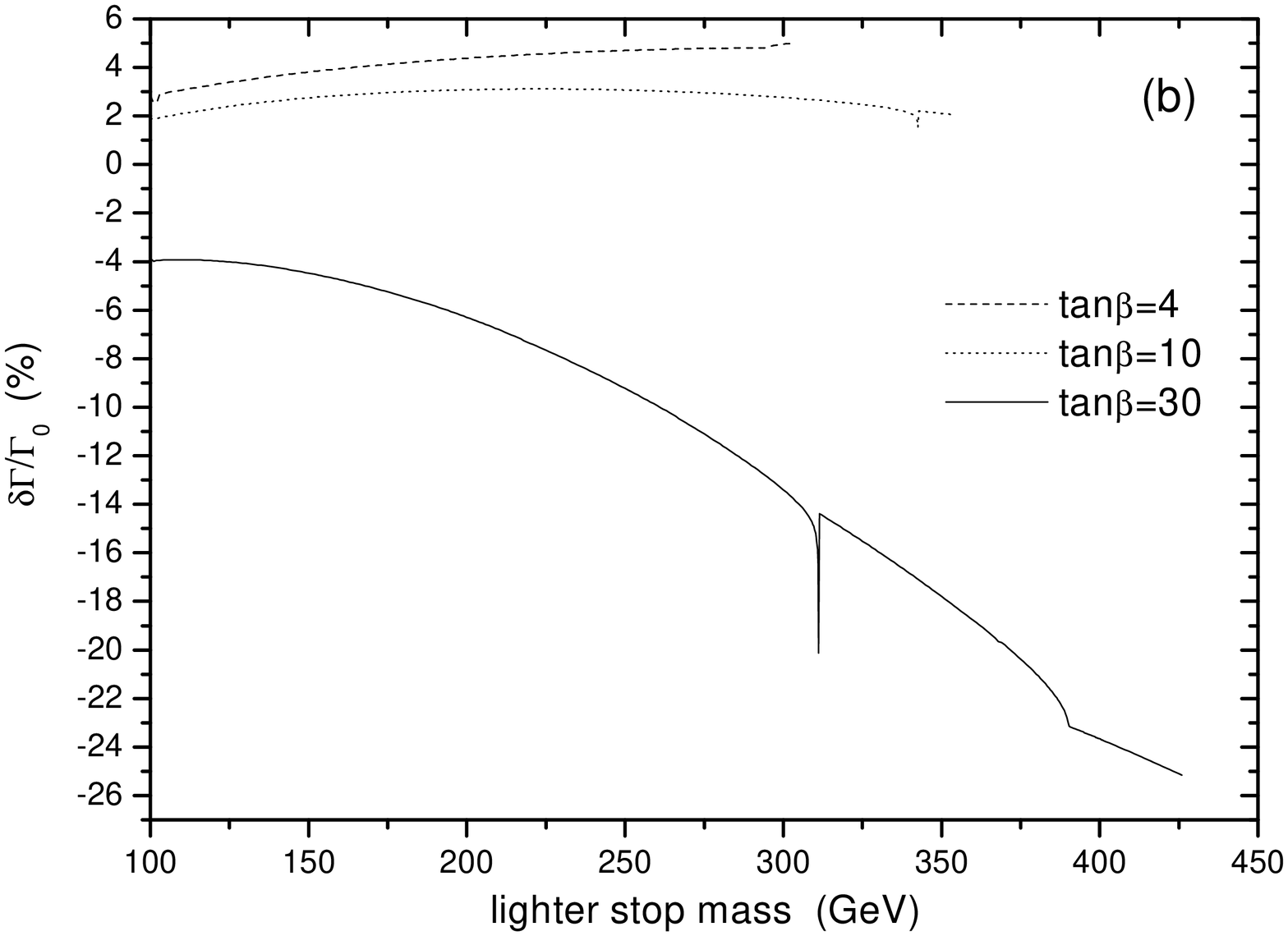, width=400pt}} \caption{The
tree-level decay width (figure $(a)$) of $\tilde{b}_2 \rightarrow
\tilde{t}_1H^-$ and its Yukawa corrections (figure $(b)$) as
functions of $m_{\tilde{t}_1}$ for $\tan\beta=4$, $10$ and $30$,
respectively, assuming $m_{A^0}=150$GeV, $\mu=M_2=400$GeV,
$A_t=A_b=1$TeV and $M_{\tilde Q}=M_{\tilde U}=M_{\tilde D}$.}
\label{mst2}
\end{figure}
\begin{figure}
\centerline{\epsfig{file=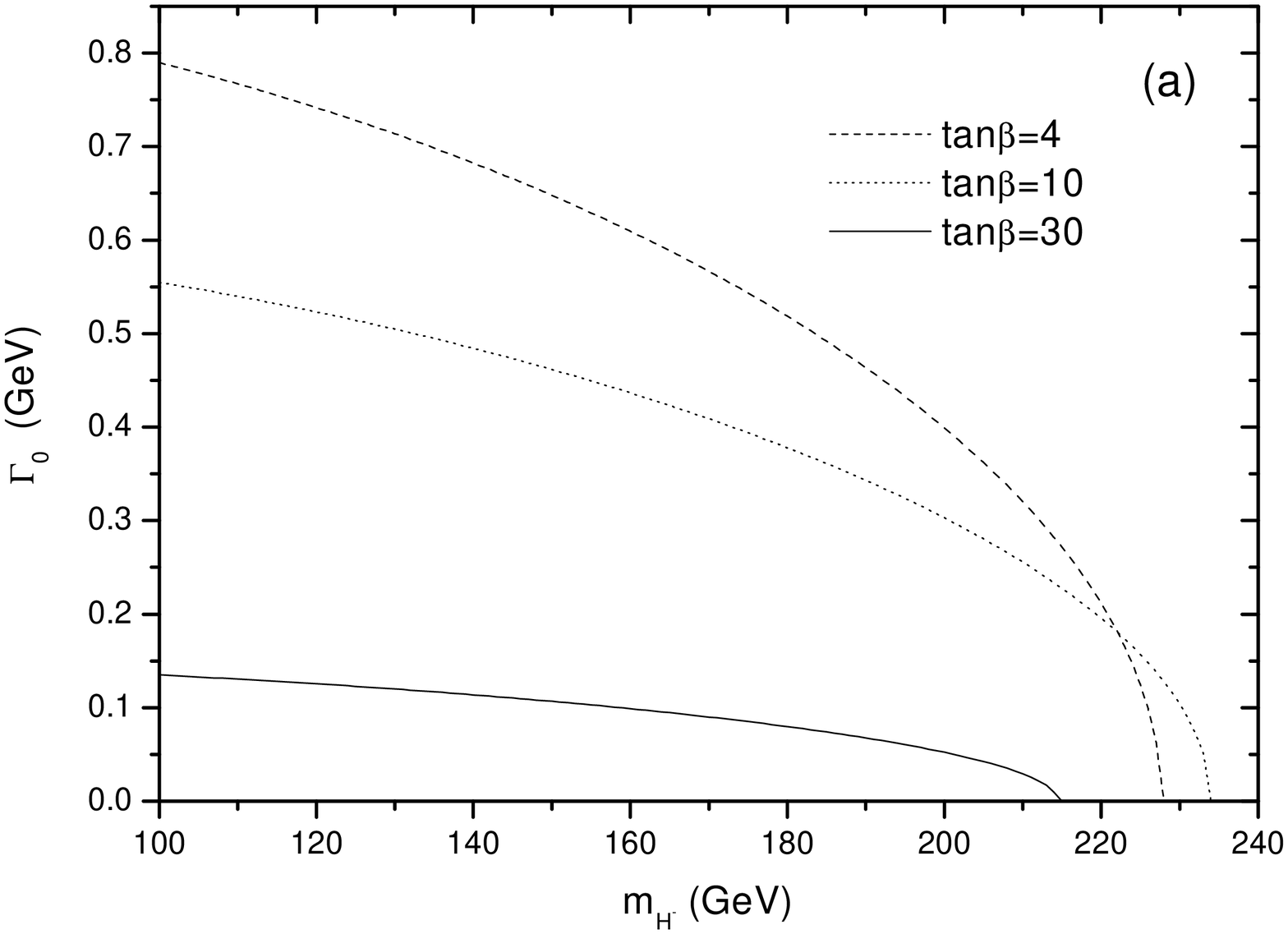, width=400pt}}
\centerline{\epsfig{file=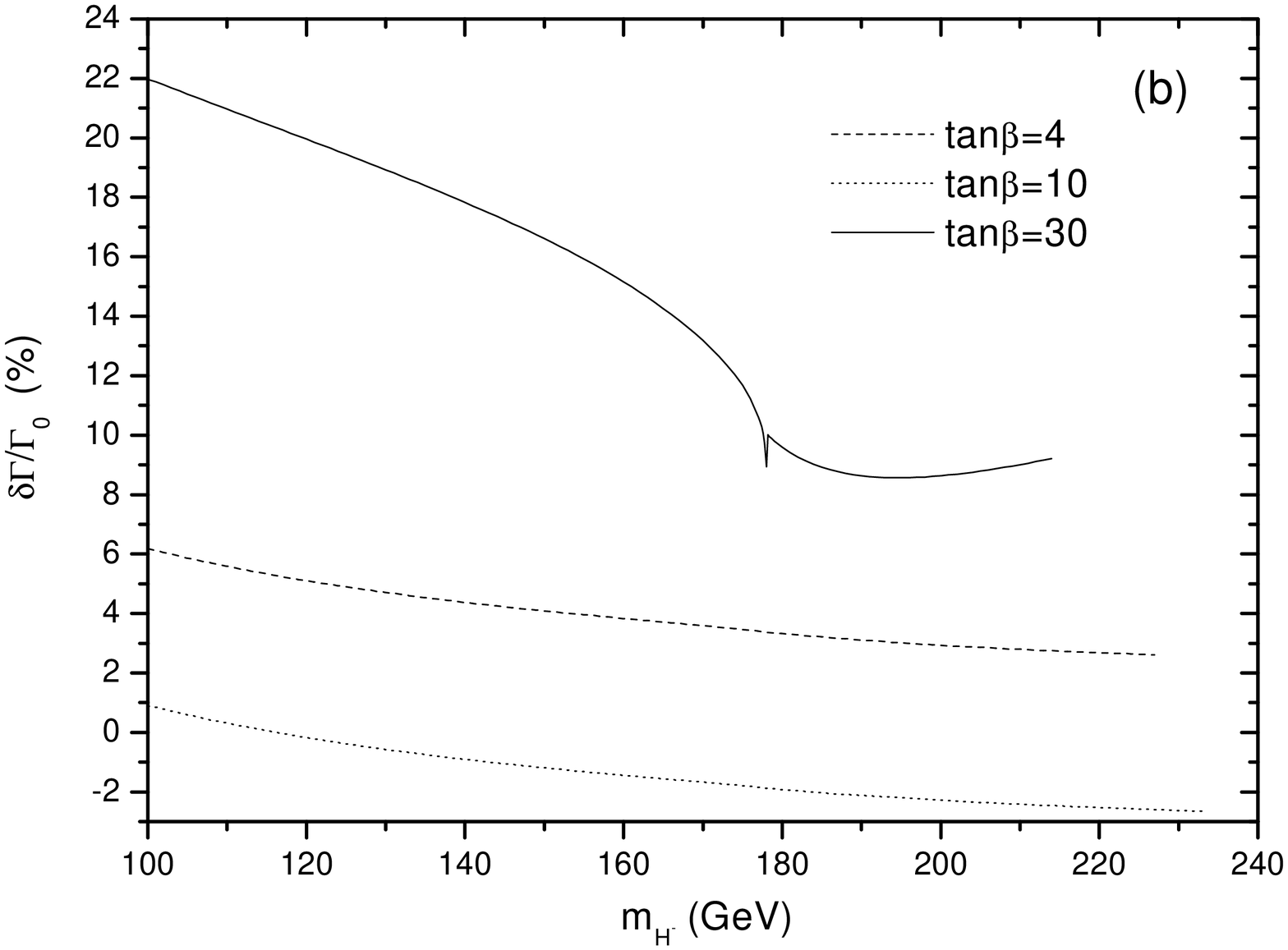, width=400pt}} \caption[]{The
tree-level decay width (figure $(a)$) of $\tilde{b}_1 \rightarrow
\tilde{t}_1H^-$ and its Yukawa corrections (figure $(b)$) as
functions of $m_{H^-}$ for $\tan\beta=4$, $10$ and $30$,
respectively, assuming $m_{\tilde{t}_1}=170$GeV, $\mu=M_2=400$GeV,
$A_t=A_b=1$TeV and $M_{\tilde Q}=M_{\tilde U}=M_{\tilde D}$.}
\label{mhp1}
\end{figure}
\begin{figure}
\centerline{\epsfig{file=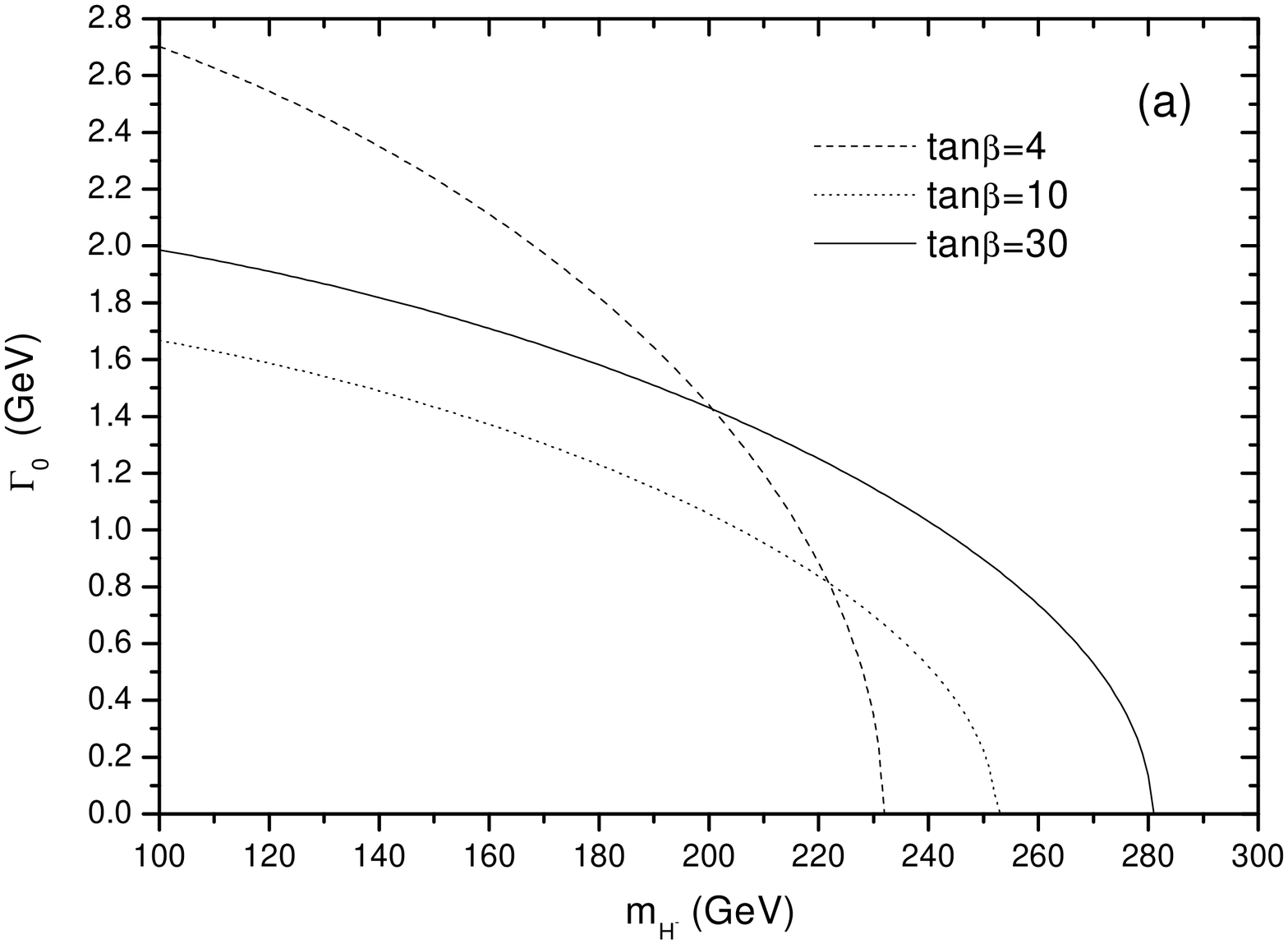, width=400pt}}
\centerline{\epsfig{file=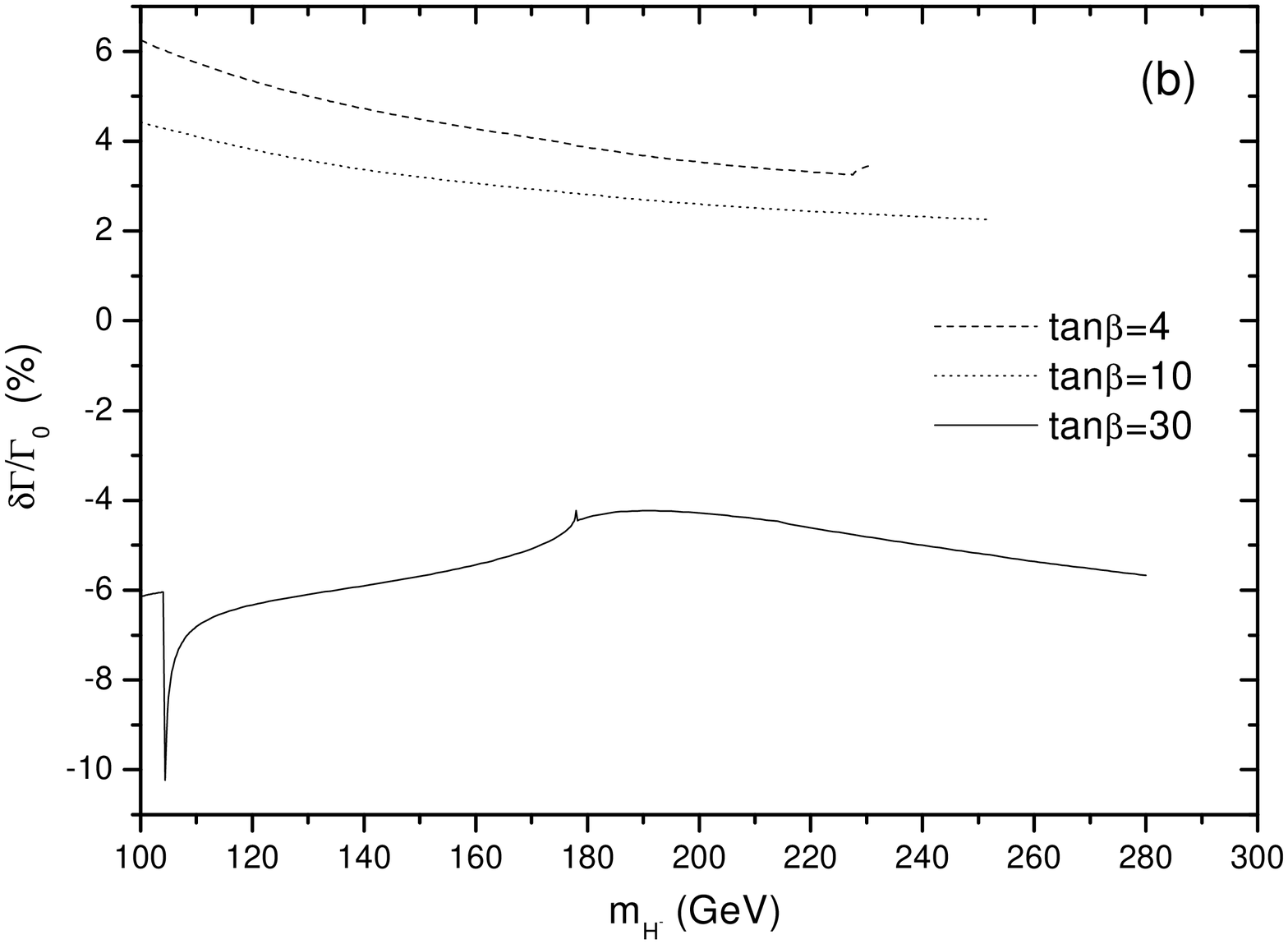, width=400pt}} \caption{The
tree-level decay width (figure $(a)$) of $\tilde{b}_2 \rightarrow
\tilde{t}_1H^-$ and its Yukawa corrections (figure $(b)$) as
functions of $m_{H^-}$ for $\tan\beta=4$, $10$ and $30$,
respectively, assuming $m_{\tilde{t}_1}=170$GeV, $\mu=M_2=400$GeV,
$A_t=A_b=1$TeV and $M_{\tilde Q}=M_{\tilde U}=M_{\tilde D}$.}
\label{mhp2}
\end{figure}
\begin{figure}
\centerline{\epsfig{file=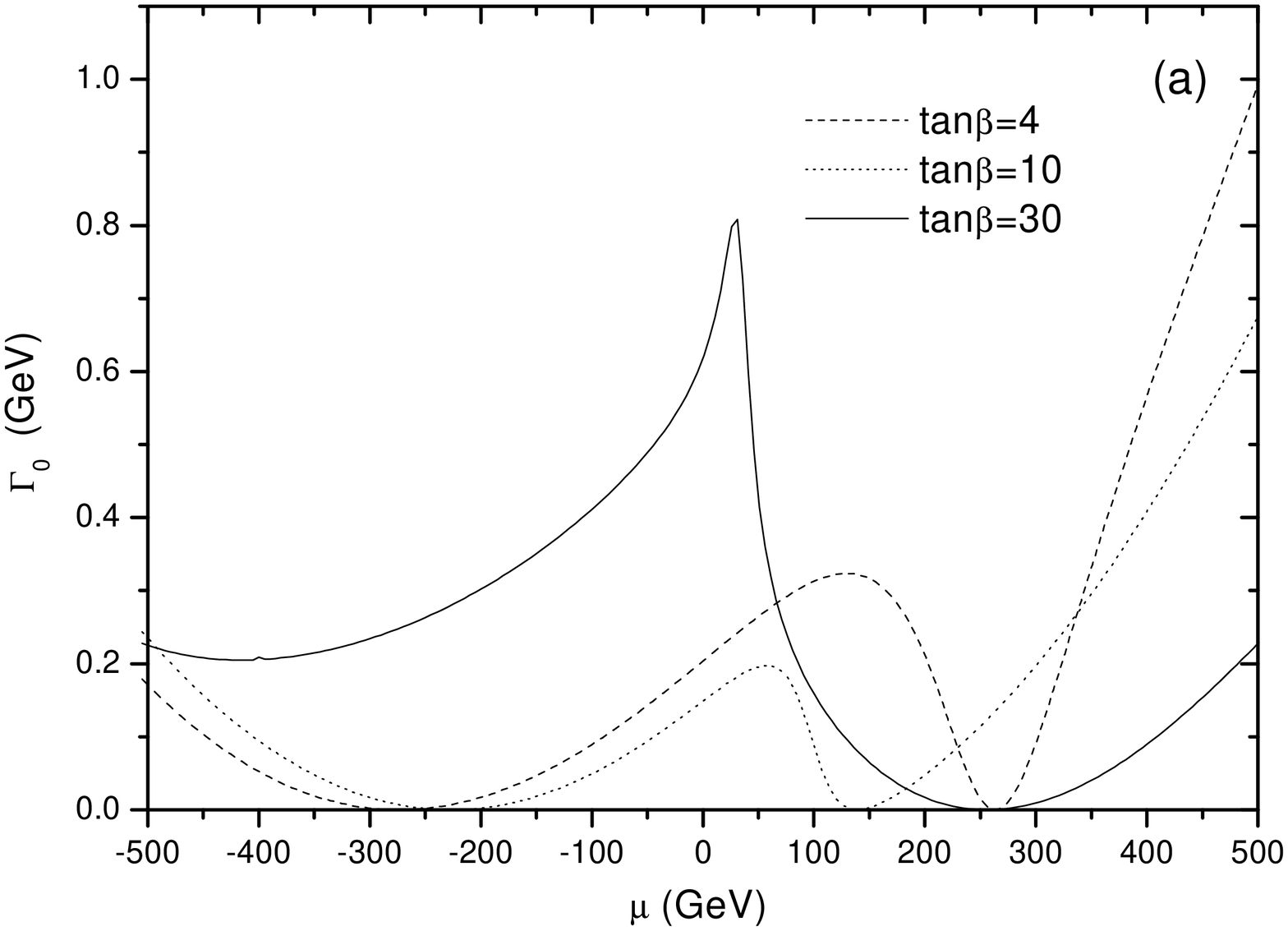, width=400pt}}
\centerline{\epsfig{file=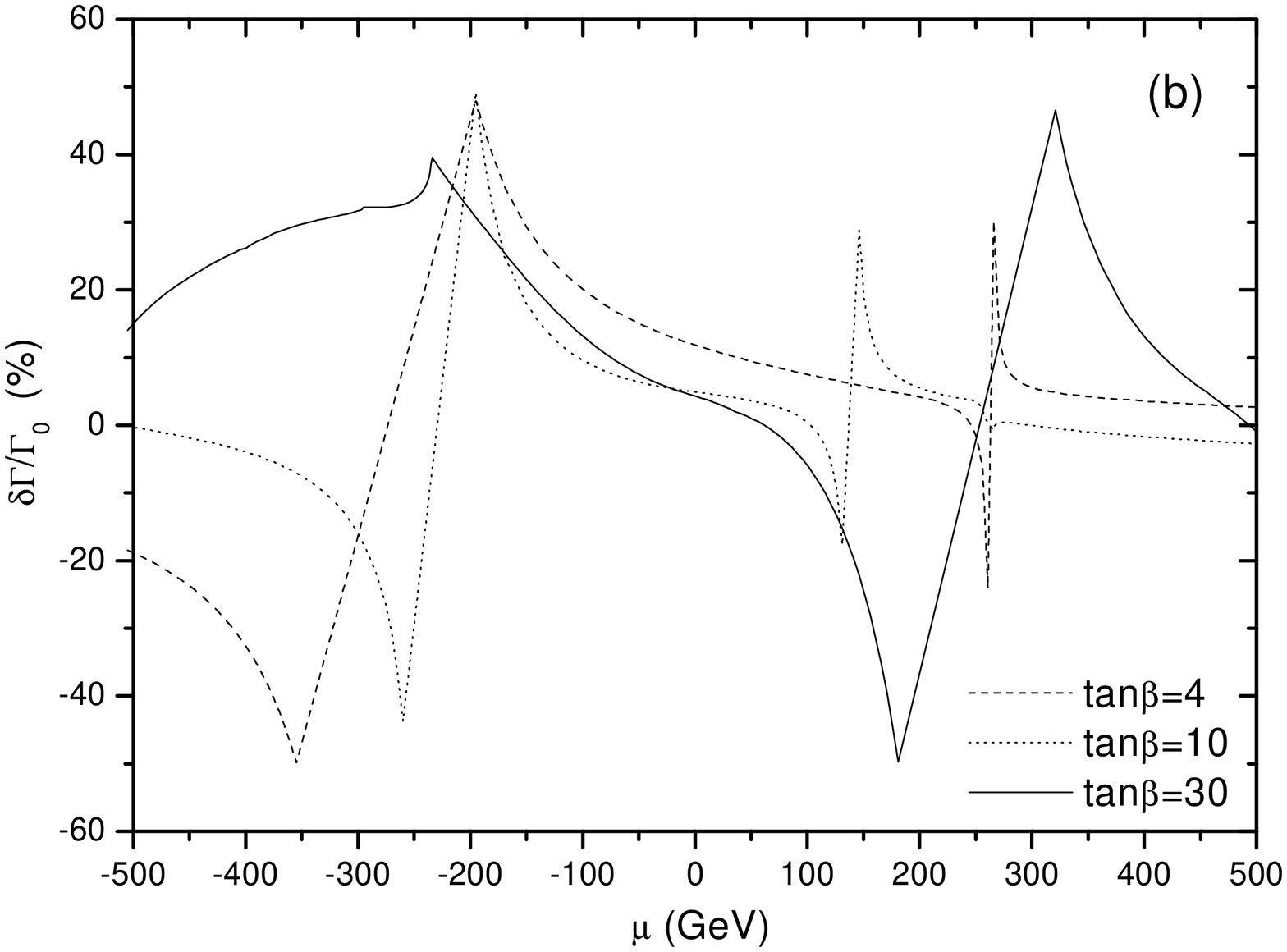, width=400pt}} \caption{The
tree-level decay width (figure $(a)$) of $\tilde{b}_1 \rightarrow
\tilde{t}_1H^-$ and its Yukawa corrections (figure $(b)$) as
functions of $\mu$ for $\tan\beta=4$, $10$ and $30$, respectively,
assuming $m_{\tilde{t}_1}=170$GeV, $M_2=400$GeV, $A_t=A_b=1$TeV,
$m_{A^0}=150$GeV and $M_{\tilde Q}=M_{\tilde U}=M_{\tilde D}$.}
\label{mu1}
\end{figure}
\begin{figure}
\centerline{\epsfig{file=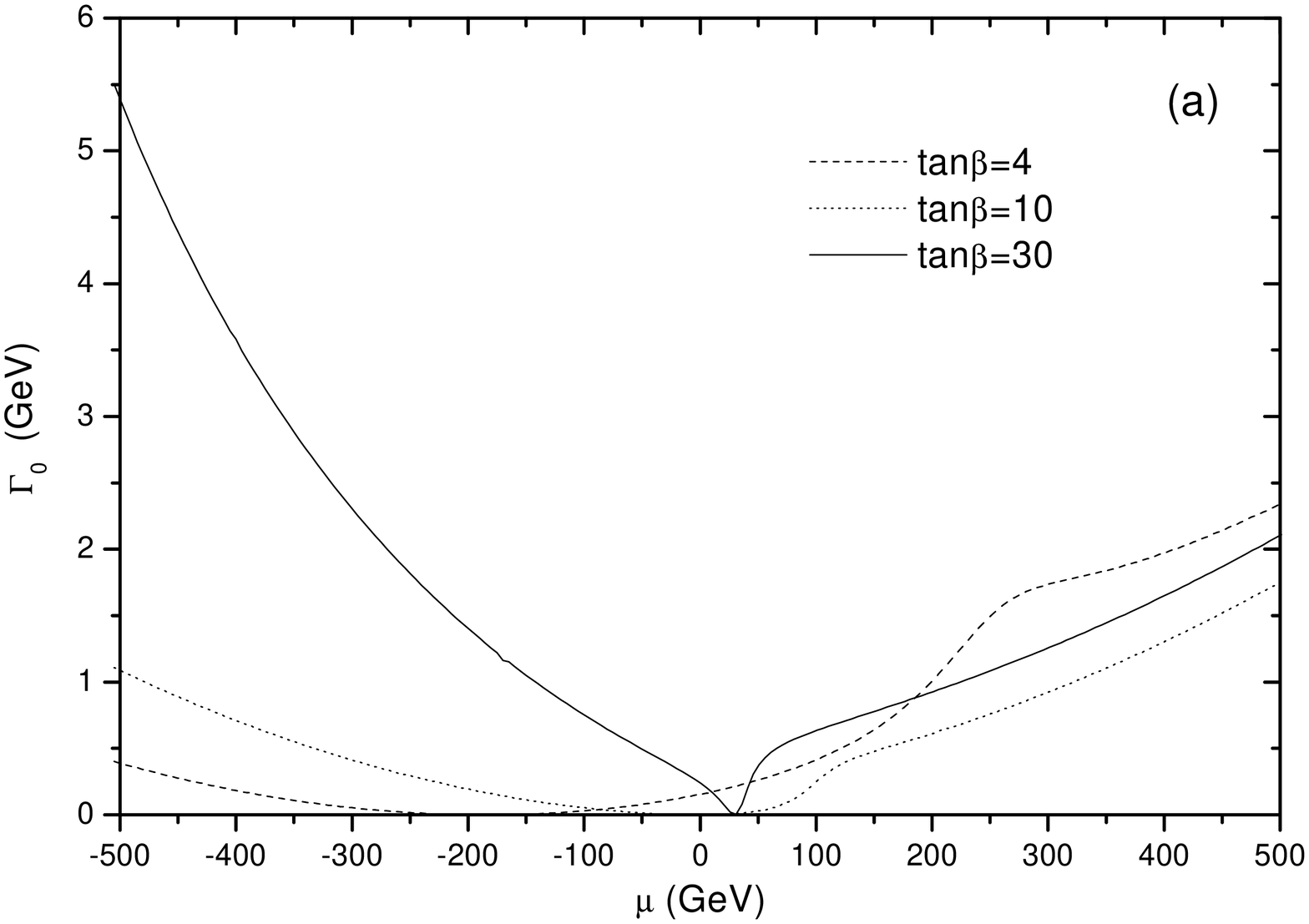, width=400pt}}
\centerline{\epsfig{file=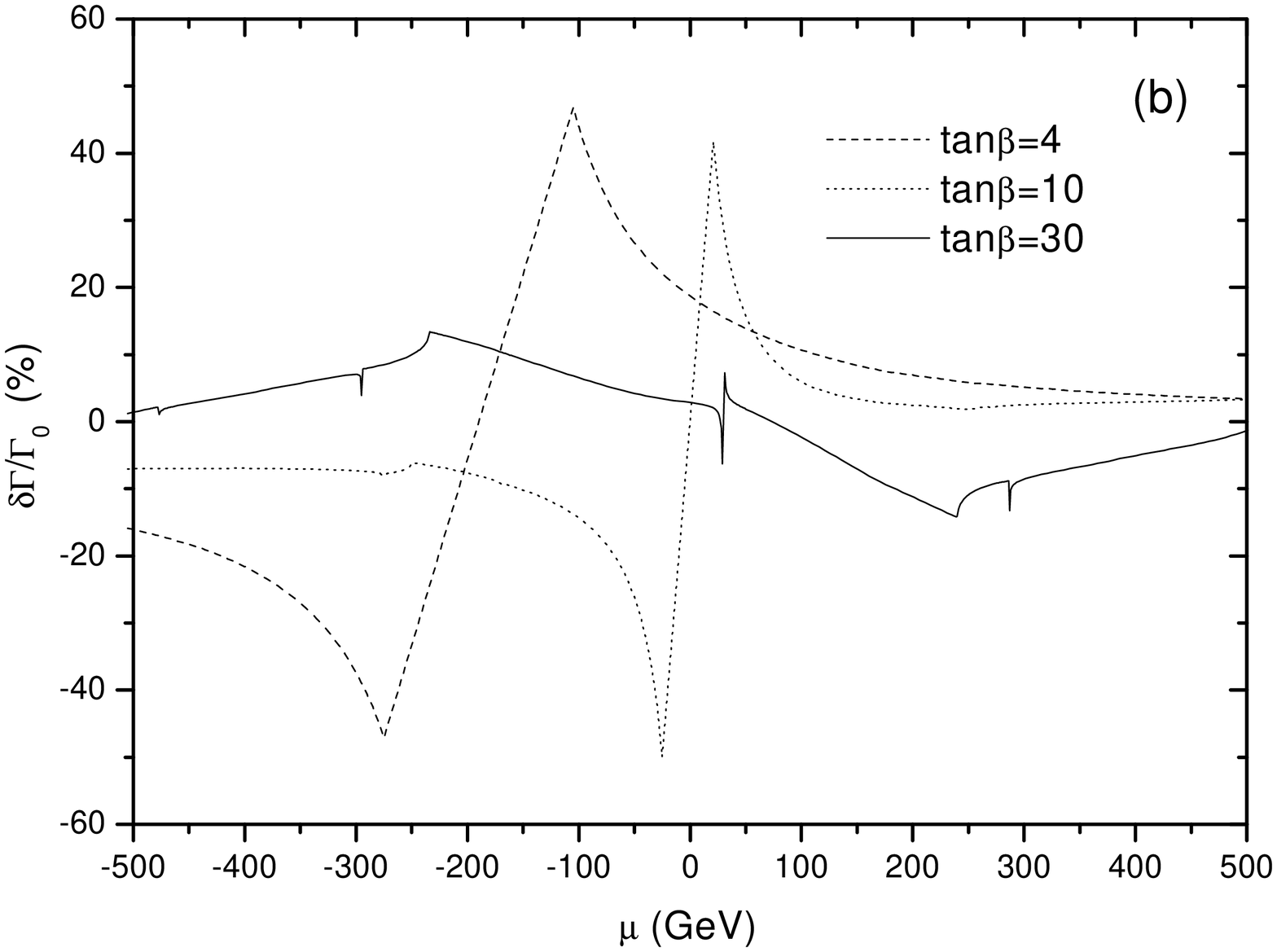, width=400pt}} \caption{The
tree-level decay width (figure $(a)$) of $\tilde{b}_2 \rightarrow
\tilde{t}_1H^-$ and its Yukawa corrections (figure $(b)$) as
functions of $\mu$ for $\tan\beta=4$, $10$ and $30$, respectively,
assuming $m_{\tilde{t}_1}=170$GeV, $M_2=400$GeV, $A_t=A_b=1$TeV,
$m_{A^0}=150$GeV and $M_{\tilde Q}=M_{\tilde U}=M_{\tilde D}$.}
\label{mu2}
\end{figure}

\begin{figure}
\centerline{\epsfig{file=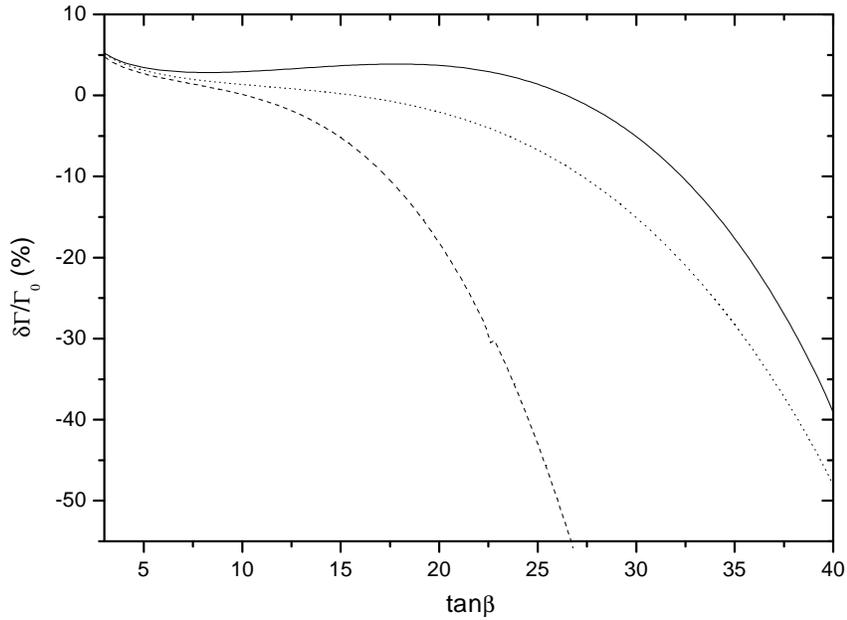, width=400pt}} \caption{The
Yukawa corrections of $\tilde{b}_2 \rightarrow \tilde{t}_1H^-$ as
a function of $\tan\beta$, assuming $m_{\tilde{t}_1}=170$GeV,
$\mu=M_2=400$GeV, $A_t=A_b=1$TeV, $m_{A^0}=150$GeV and $M_{\tilde
Q}=M_{\tilde U}=M_{\tilde D}$. The dashed line corresponds to the
corrections using the on-shell bottom quark mass, the dotted line
to the improved result only using the QCD running mass $m_b(Q)$,
and the solid line to the improved result using the replacement
Eq.(\ref{replacement}).} \label{tan}
\end{figure}

\end{document}